\pdfoutput=1
\documentclass[aps,
amsmath,amssymb,
reprint,
superscriptaddress
]{revtex4-1}
\usepackage[unicode]{hyperref}

\usepackage{graphicx}
\usepackage{color}
\usepackage{amsmath}
\usepackage{amssymb}
\usepackage{physics}
\usepackage{amsfonts}
\usepackage{colortbl}
\usepackage{xcolor}
\usepackage{dcolumn}% Align table columns on decimal point
\usepackage{bm}% bold math
\usepackage[utf8]{inputenc}
\usepackage[T1]{fontenc}
\usepackage{mathptmx}
\usepackage{etoolbox}
\usepackage{booktabs}
\usepackage{placeins}

\usepackage{lineno}

\newif\ifrevred \revredfalse
\newcommand{\rev}[1]{\ifrevred\textcolor{red}{#1}\else#1\fi}
\newcommand{\revcolor}{\ifrevred\color{red}\fi}
\colorlet{RED}{red}

\makeatletter
\def\@email#1#2{%
 \endgroup
 \patchcmd{\titleblock@produce}
  {\frontmatter@RRAPformat}
  {\frontmatter@RRAPformat{\produce@RRAP{*#1\href{mailto:#2}{#2}}}\frontmatter@RRAPformat}
  {}{}
}%
\makeatother
\begin{document}

% \preprint{AIP/123-QED}

\title{Direct measurement of the on-shot focal-spot size of a kilojoule-class petawatt laser}
% Force line breaks with \\
\author{Ryunosuke Takizawa}
 \email{takizawa.ryunosuke.ile@osaka-u.ac.jp}
 \affiliation{Institute of Laser Engineering, The University of Osaka, 2-6 Yamada-Oka, Suita, Osaka 565-0871, Japan}
\author{Keisuke Takahashi}
 \affiliation{Institute of Laser Engineering, The University of Osaka, 2-6 Yamada-Oka, Suita, Osaka 565-0871, Japan}
\author{Yuhei Kawakami}
 \affiliation{Institute of Laser Engineering, The University of Osaka, 2-6 Yamada-Oka, Suita, Osaka 565-0871, Japan}
\author{Hiroki Matsubara}
 \affiliation{Department of Physics, Graduate School of Science, The University of Osaka, Toyonaka, 560-0043, Japan}
 \affiliation{Institute of Laser Engineering, The University of Osaka, 2-6 Yamada-Oka, Suita, Osaka 565-0871, Japan}
\author{Yuga Karaki}
 \affiliation{Department of Physics, Graduate School of Science, The University of Osaka, Toyonaka, 560-0043, Japan}
 \affiliation{Institute of Laser Engineering, The University of Osaka, 2-6 Yamada-Oka, Suita, Osaka 565-0871, Japan}
\author{Rinya Akematsu}
 \affiliation{Department of Physics, Graduate School of Science, The University of Osaka, Toyonaka, 560-0043, Japan}
 \affiliation{Institute of Laser Engineering, The University of Osaka, 2-6 Yamada-Oka, Suita, Osaka 565-0871, Japan}
\author{Ryou Oomura}
 \affiliation{Department of Physics, Graduate School of Science, The University of Osaka, Toyonaka, 560-0043, Japan}
 \affiliation{Institute of Laser Engineering, The University of Osaka, 2-6 Yamada-Oka, Suita, Osaka 565-0871, Japan}
\author{Kai Kimura}
 \affiliation{Department of Physics, Graduate School of Science, The University of Osaka, Toyonaka, 560-0043, Japan}
 \affiliation{Institute of Laser Engineering, The University of Osaka, 2-6 Yamada-Oka, Suita, Osaka 565-0871, Japan}
 \author{Yuki Tamaru}%
 \affiliation{Institute of Laser Engineering, The University of Osaka, 2-6 Yamada-Oka, Suita, Osaka 565-0871, Japan}
 \author{Yasunobu Arikawa}%
 \affiliation{Institute of Laser Engineering, The University of Osaka, 2-6 Yamada-Oka, Suita, Osaka 565-0871, Japan}
 \author{Koji Tsubakimoto}%
 \affiliation{Institute of Laser Engineering, The University of Osaka, 2-6 Yamada-Oka, Suita, Osaka 565-0871, Japan}
\author{Shinsuke Fujioka}%
 \affiliation{Institute of Laser Engineering, The University of Osaka, 2-6 Yamada-Oka, Suita, Osaka 565-0871, Japan}
  \affiliation{National Institute for Fusion Science, 322-6 Oroshi, Toki, Gifu 509-5292, Japan}

% \affiliation{
% Authors' institution and/or address%\\This line break forced with \textbackslash\textbackslash
% }%

% \author{C. Author}
%  % \homepage{http://www.Second.institution.edu/~Charlie.Author.}
% \affiliation{%
% Second institution and/or address%\\This line break forced% with \\
% }%

\date{\today}% It is always \today, today,

\begin{abstract}
We report a direct, on-shot determination of the focal-spot size of a kilojoule-class, four-beam petawatt laser.
Encircled-energy curves were inferred from the energy transmitted around aluminum-coated spheres of varying diameters placed at the focal plane.
\rev{The 50\% encircled-energy diameter $D_{50}$ of each beam was obtained by fitting a two-component Gaussian mixture with Bayesian inference.}
\rev{With all deformable mirrors optimized in closed loop, the best-corrected alignment beam reached $D_{50} = 34^{+4}_{-5}~\mu\mathrm{m}$, three times the diffraction limit of $10~\mu\mathrm{m}$.}
\rev{Under full-power conditions, the thermal distortion of the main amplifier broadened this beam to $D_{50} = 102^{+9}_{-8}~\mu\mathrm{m}$ and reduced its peak intensity ninefold, to $2.2^{+0.6}_{-0.9}\times10^{18}~\mathrm{W/cm^2}$ at 250~J and 1.3~ps.}
\rev{The energy fraction of the narrow component decreased from 68\% to 38\%.}
\rev{These reference values serve facility users, and a deformable mirror upstream of the pulse compressor, now on one beam only, doubles the on-shot peak intensity.}
\end{abstract}



\maketitle %\maketitle must follow title, authors, abstract and \pacs

\section{Introduction}

The invention of chirped-pulse amplification\cite{Strickland1985-bo} enabled the development of ultrashort-pulse lasers with sub-picosecond durations.
Furthermore, combining chirped-pulse amplification with optical parametric amplification has made it possible to generate higher-energy short-pulse lasers\cite{Perry1999-ot}.
In particular, systems based on Nd:glass lasers can deliver kilojoule-class pulse energies; even with relatively long pulse durations on the order of picoseconds, intensities exceeding $10^{18}$ W/cm$^2$ can be achieved.
These kilojoule-class petawatt lasers\cite{Danson2019-fg} remain scarce worldwide.
Representative facilities include LFEX\cite{Miyanaga2006-bq}, OMEGA-EP\cite{Waxer2008-sn}, NIF-ARC\cite{Barty2004-ib}, LMJ-PETAL\cite{Blanchot2017-xv}, ELI-NP\cite{Radier2022-ug}, and ShenGuang-II\cite{Zhu2018-zy}.
These large-scale laser facilities actively support international collaborations and are commonly accessible to external users.
However, because the systems are complex and highly specialized, it is difficult for users to fully grasp all system characteristics.
Beam tuning and operation therefore rely on experienced technical staff, and within limited machine time it is not straightforward to measure on-shot laser parameters in detail.
Against this backdrop, systematically reporting each facility's optical configuration and laser specifications is critically important for cross-comparison among experiments and for reproducibility.

The laser focal intensity is one of the fundamental parameters governing laser--plasma interactions, and many physical processes follow intensity-scaling laws.
For example, the maximum ion energy in target-normal sheath acceleration\cite{Fuchs2006-zs}, the mean electron energy associated with $J\times B$ (ponderomotive) acceleration\cite{Wilks1992-yp}, and the laser absorption fraction\cite{Davies2008-ux} all depend strongly on the focal intensity.
Accordingly, accurate knowledge of the on-shot focal-spot size is indispensable for the physical interpretation of experimental results.

In this paper, we quantitatively evaluate the focal-spot characteristics of the LFEX laser at The University of Osaka and clarify its current focusing performance.
For single-shot laser systems, direct observation of the on-shot focal spot is generally difficult, \rev{and each facility has developed its own approach}.
\rev{The focal-spot diagnostic of OMEGA-EP measures the wavefront of a sampled beam on shot and reconstructs the far-field profile from it\cite{Bromage2008-as}.}
\rev{PETAL images the second- and third-harmonic emission from the target surface on shot\cite{Raffestin2019-pt}.}
In contrast, the wavefront at the final optical stage of LFEX has not been measured, and on-shot focal-spot information is therefore unavailable to users.
This uncertainty in the focal-spot profile directly leads to ambiguity in the on-target intensity, which can critically affect the physical interpretation of experimental data.
To address this, we directly measured the focal spot under shot conditions and provide reference spot-size data that users can consult.

%%%%%%%%%%%%%%%%%%%%%%%%%%%%%%%%%%%%%%%%%%%%%%%%%%%%%%%%%%%%%%%%%%%%%%%%%%%%%%%%%%%%%

\section{LFEX laser system}
\label{sec:exp_conf}

At the final stage of LFEX, four beams with square cross sections of $400~\mathrm{mm}$ on a side are arranged in a $2\times2$ array.
In the laser amplifier chain, the beams are designated as follows: the upper right is H1, the upper left is H2, the lower right is H3, and the lower left is H4, as viewed from the downstream side of the laser system.
The final focusing optic is a 1-m-class off-axis parabolic mirror operated at an incidence angle of $60^\circ$, which simultaneously focuses all four beams.
The mirror has a reflective focal length of \rev{$f = 4000~\mathrm{mm}$}.
\rev{The diffraction-limited far-field pattern of a uniformly illuminated square aperture of side $D = 400~\mathrm{mm}$ at $\lambda = 1.053~\mu\mathrm{m}$ has a central lobe of full width $2\lambda f/D = 21~\mu\mathrm{m}$, which contains 82\% of the energy.}
\rev{The diameter of the circle that encloses 50\% of the energy, which is the spot-size metric used in this paper, is $D_{50}^{\mathrm{DL}} = 0.96\,\lambda f/D = 10~\mu\mathrm{m}$ for the square aperture.}

The preamplifier chain consists of a four-pass rod amplifier followed by a two-stage rod amplifier; the beam is divided into four between these modules.
\rev{A seed of about $0.5~\mathrm{mJ}$} from the upstream optical parametric chirped-pulse amplification system is amplified in the preamplifier to $5$--$10~\mathrm{J}$ per beam.
The main amplifier section comprises eight four-pass disk-amplifier stages (DA400), with spatial filters inserted after the second and fourth passes.
The beam entering the main amplifier is expanded to a square aperture of $360~\mathrm{mm}$ on a side, and the amplified energy is \rev{$400$--$500~\mathrm{J}$} per beam.

In the disk amplifiers, heating of the laser glass by the flashlamp pump leads to thermal expansion and deformation, causing substantial wavefront aberration of the transmitted beam.
LFEX employs deformable mirrors (DFMs) to compensate these aberrations.
Two DFMs, DFM75 and DFM125, are installed in the main amplifier: DFM75, placed at the entrance-mirror location of the main amplifier, compensates wavefront errors originating from the first and second passes, while DFM125, positioned at the entrance-mirror location for the third pass, compensates thermal distortions associated with the third and fourth passes.
In addition, only for the H2 beam, a DFM (DFM400) is inserted as the last mirror immediately upstream of the pulse compressor to correct optical wavefront errors introduced by the compressor.
Consequently, the H2 beam exhibits superior focusing performance compared with the other beams.

The basic correction procedure for DFM75, DFM125, and DFM400 is based on reproducing a point-source image.
For DFM75 and DFM125, a pinhole with a diameter smaller than the diffraction-limited spot is placed at the focal planes of the spatial filters located after the second and fourth passes, respectively; for DFM400, the pinhole is placed at the focus of the off-axis parabolic mirror.
The light transmitted through the pinhole is measured with a wavefront sensor and used as the reference.
DFM75 and DFM125 are tuned using shots under conditions equivalent to full-power operation until the measured wavefront approaches the reference data.
Because DFM400 targets only optics-induced aberrations, it is operated in a closed loop via feedback using the alignment laser.

During alignment, the system's \rev{alignment laser, a 6-Hz pulsed beam,} is used to optimize the DFMs prior to a shot.
In contrast, actual shots employ broadband nanosecond pulses delivered by the optical parametric chirped-pulse amplification system.
The alignment beam and the seed pulse are matched upstream and propagated coaxially through the DFMs and the main amplifier.
Further details of the LFEX laser system are provided in Ref.~\cite{Miyanaga2006-bq}.

%%%%%%%%%%%%%%%%%%%%%%%%%%%%%%%%%%%%%%%%%%%%%%%%%%%%%%%%%%%%%%%%%%%%%%%%%%%%%

\section{Focal-spot measurement}
\label{sec:exp_ana}

\begin{figure}[!htbp]
  \centering
  \includegraphics[width=0.8\linewidth]{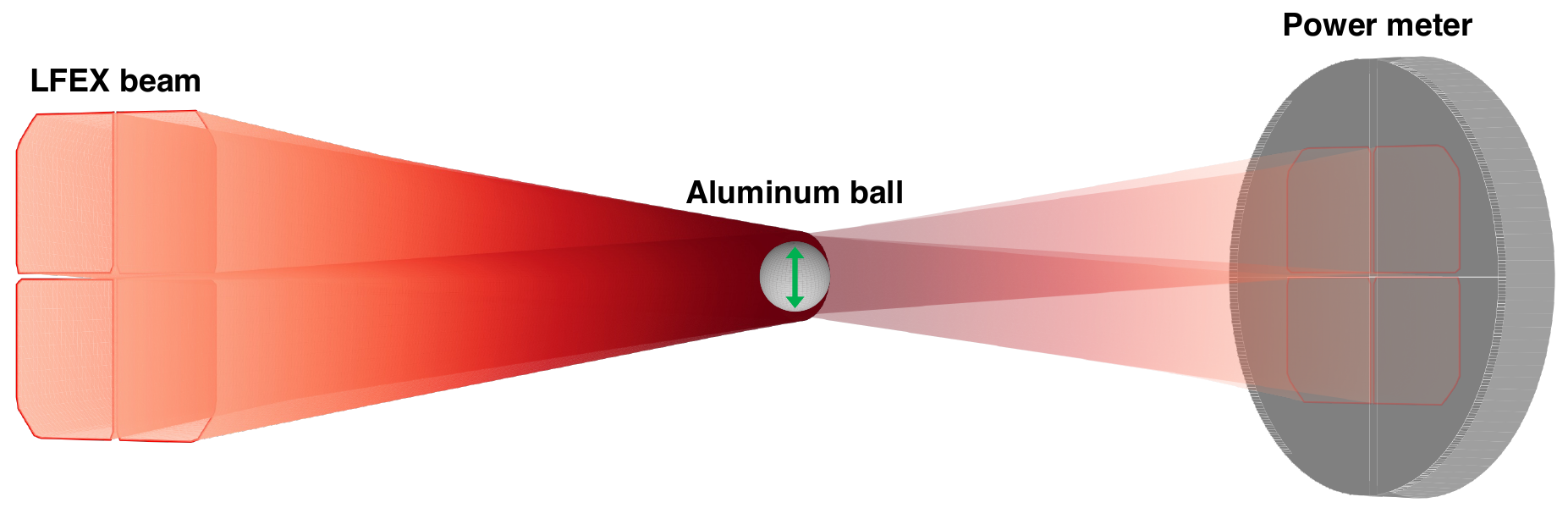}
  \caption{
  \textbf{Configuration for the focal-spot measurement \rev{(schematic)}.}
  \rev{An aluminum-coated glass sphere was placed at the target position, and the laser energy transmitted around the sphere was measured outside the target chamber with a calorimeter while varying the sphere diameter.
  Although the drawing shows all four beams, only one beam was fired on each shot.}
  }
  \label{fig:exp_setup}
\end{figure}

Aluminum-coated glass spheres \rev{with diameters of 25--300~$\mu$m, measured on microscope images,} were fixed in place using microfiber and placed at the laser focal plane.
The aluminum coating prevents shine-through, i.e., transmission of the laser light through the transparent glass sphere before an overdense plasma forms on its surface, ensuring that each sphere blocks the light within its geometrical cross section.
The spot size was evaluated from the laser energy that passed around the spheres.
\rev{The light passing around each sphere diverged beyond the focus, left the target chamber through a vacuum window, and was measured with a calorimeter positioned downstream along the beam axis (Fig.~\ref{fig:exp_setup}).}
\rev{The calorimeter (Scientech 380801 head with an S310 readout) has a calibration uncertainty of $\pm5\%$.}
\rev{One beam was fired per shot, and the four beams were measured in separate series of shots.}

The main amplifier was operated at its maximum charging voltage to reproduce the wavefront of full-power shots.
\rev{The seed energy was reduced to keep the transmitted energy within the dynamic range of the calorimeter.}
The transmitted energy of each shot was normalized by the beam energy monitored on the same shot \rev{before pulse compression with the incident-beam monitoring and analyzing package (IMAP)}, and reference shots without a sphere defined unity transmittance.
\rev{Table~\ref{tab:shots} lists all full-power shots; one shot was taken per sphere diameter, and the beam energy before compression ranged from 2.7 to 13~J because the seed energy was raised for the larger spheres.}
\rev{For comparison, the spot of the alignment laser was measured in the same way, but with a near-field camera triggered on single pulses of the 6-Hz laser instead of the calorimeter.}
\rev{The transmittance is the ratio of the integrated brightness of the images with and without the sphere, each after subtraction of the background integrated over the same area.}
\rev{The alignment-laser measurement was made at a separate position that simulates the target position, with the beam focused by the same off-axis parabolic mirror.}
\rev{The alignment laser propagates through the unpumped main amplifier and therefore does not experience its thermal wavefront distortion.}

\begin{table*}[!tbp]
\centering
{\revcolor
\begin{tabular}{lccccc}
\toprule
Beam & Shot & Sphere diameter ($\mu$m) & Beam energy (J) & Transmitted energy (J) & Transmittance \\
\midrule
H1 & 5692 & -- & $2.7 \pm 0.3$ & $1.74 \pm 0.09$ & $1.00 \pm 0.07$ \\
 & 5694 & $110 \pm 11$ & $4.7 \pm 0.5$ & $1.95 \pm 0.10$ & $0.63 \pm 0.06$ \\
 & 5695 & $214 \pm 21$ & $7.8 \pm 0.8$ & $0.64 \pm 0.03$ & $0.13 \pm 0.05$ \\
\midrule
H2 & 5727 & -- & $2.8 \pm 0.3$ & $1.75 \pm 0.09$ & $1.00 \pm 0.07$ \\
 & 5774 & $34 \pm 3$ & $3.0 \pm 0.3$ & $1.74 \pm 0.09$ & $0.94 \pm 0.07$ \\
 & 5730 & $56 \pm 6$ & $2.7 \pm 0.3$ & $1.18 \pm 0.06$ & $0.70 \pm 0.06$ \\
 & 5726 & $100 \pm 10$ & $4.4 \pm 0.4$ & $1.37 \pm 0.07$ & $0.50 \pm 0.06$ \\
 & 5775 & $106 \pm 11$ & $3.7 \pm 0.4$ & $0.98 \pm 0.05$ & $0.42 \pm 0.05$ \\
 & 5729 & $163 \pm 16$ & $7.0 \pm 0.7$ & $1.45 \pm 0.07$ & $0.33 \pm 0.05$ \\
 & 5731 & $261 \pm 26$ & $12.9 \pm 1.3$ & $0.60 \pm 0.03$ & $0.07 \pm 0.05$ \\
\midrule
H3 & 5646 & -- & $2.7 \pm 0.3$ & $1.57 \pm 0.08$ & $1.03 \pm 0.07$ \\
 & 5687 & -- & $3.7 \pm 0.4$ & $2.05 \pm 0.10$ & $0.97 \pm 0.07$ \\
 & 5696 & $47 \pm 5$ & $3.3 \pm 0.3$ & $1.64 \pm 0.08$ & $0.88 \pm 0.07$ \\
 & 5648 & $94 \pm 9$ & $5.8 \pm 0.6$ & $2.05 \pm 0.10$ & $0.63 \pm 0.06$ \\
 & 5650 & $101 \pm 10$ & $5.0 \pm 0.5$ & $1.64 \pm 0.08$ & $0.58 \pm 0.06$ \\
 & 5651 & $198 \pm 20$ & $9.5 \pm 1.0$ & $0.84 \pm 0.04$ & $0.16 \pm 0.05$ \\
\midrule
H4 & 5701 & -- & $3.0 \pm 0.3$ & $1.73 \pm 0.09$ & $1.00 \pm 0.07$ \\
 & 5698 & $110 \pm 11$ & $4.5 \pm 0.4$ & $2.00 \pm 0.10$ & $0.79 \pm 0.06$ \\
 & 5700 & $158 \pm 16$ & $4.6 \pm 0.5$ & $1.27 \pm 0.06$ & $0.48 \pm 0.06$ \\
 & 5699 & $204 \pm 20$ & $6.1 \pm 0.6$ & $1.11 \pm 0.06$ & $0.32 \pm 0.05$ \\
\bottomrule
\end{tabular}}
\caption{\rev{Shot list of the full-power measurement. The sphere diameter was measured on microscope images (uncertainty 10\%), the beam energy was monitored before pulse compression with IMAP (uncertainty 10\%), and the transmitted energy was measured with the calorimeter (uncertainty 5\%). The transmittance is the transmitted energy divided by the beam energy and normalized to the mean of the reference shots without a sphere (--) of the same beam, so that the absolute calibrations of the calorimeter and of IMAP cancel. The uncertainty of the transmittance, 5\% (relative) combined in quadrature with 0.05 (absolute), is the value used in the fit.}}
\label{tab:shots}
\end{table*}

\rev{The thermal distortion is produced by the flashlamp pumping, which was the same for all shots, and not by the extracted energy.}
\rev{Gain saturation and the nonlinear phase are weak even at full energy.}
\rev{The output fluence of 0.3--0.4~J/cm$^2$ at 400--500~J is an order of magnitude below the saturation fluence of Nd:phosphate laser glass (about 5~J/cm$^2$), and the B integral of the disk amplifier is about 0.1~rad at 500~J.}
\rev{The spot measured at reduced seed energy therefore represents that of a full-energy shot.}

\begin{figure}[!htbp]
  \centering
  \includegraphics[width=0.71\linewidth]{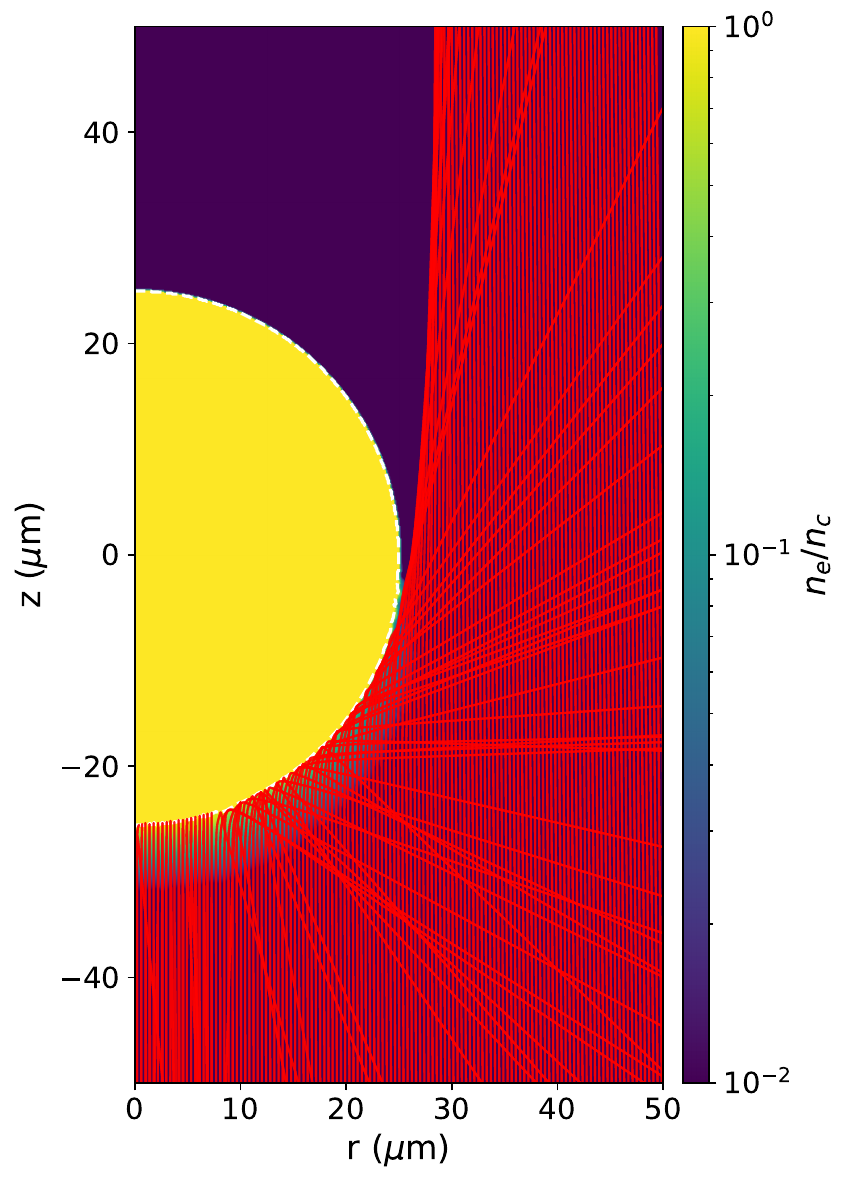}
  \caption{
  \textbf{Simulation of plasma expansion from the sphere.}
  Two-dimensional axisymmetric radiation-hydrodynamics simulation using the \texttt{FLASH} code for a $50~\mu\mathrm{m}$-diameter aluminum sphere, irradiated with the measured laser waveform scaled to the incident energy.
  The electron-density distribution at the time of peak laser intensity and the ray-tracing results, including refraction in the expanding plasma, are shown.
  The plasma expansion contributing to laser reflection extends at most \rev{about} $5~\mu\mathrm{m}$ from the original sphere surface.
  }
  \label{fig:flash}
\end{figure}

Laser irradiation ionizes and ablates the sphere, and the resulting plasma expansion increases the effective blocking radius, thereby introducing a systematic error in the measurement.
Preplasma formation driven by light arriving ahead of the main pulse (the prepulse pedestal and the foot of the pulse) could enhance this expansion.
In the measured temporal profile of the LFEX pulse, the pedestal intensity is lower than the peak intensity by more than eight orders of magnitude, and the fluence delivered earlier than $30~\mathrm{ps}$ before the peak amounts to less than $10^{-4}$ of the total pulse energy; preplasma formation prior to the main pulse is therefore negligible.
To quantitatively evaluate the expansion during the pulse, we performed a two-dimensional axisymmetric radiation-hydrodynamics simulation using the \texttt{FLASH}~\cite{Fryxell2000-om} code.
The simulation employed an aluminum sphere of radius $25~\mu\mathrm{m}$ irradiated with the measured temporal profile, including the foot of the pulse from $30~\mathrm{ps}$ before the peak, with the amplitude scaled to match the incident energy.
Figure~\ref{fig:flash} shows the electron-density distribution at the time of peak laser intensity, together with the ray-tracing results, which account for refraction in the plasma.
The rays are reflected near the critical-density surface, which remains within \rev{about} $5~\mu\mathrm{m}$ of the original sphere surface, whereas rays passing farther out traverse only the underdense corona and are only weakly deflected.
The expansion of the reflecting surface remains at this level up to the end of the pulse, by which time almost all of the pulse energy has been delivered.
The measurement uncertainty of the sphere radius is approximately 10\%.

\begin{figure*}[!tbp]
  \centering
  \includegraphics[width=0.75\textwidth,height=0.60\textheight,keepaspectratio]{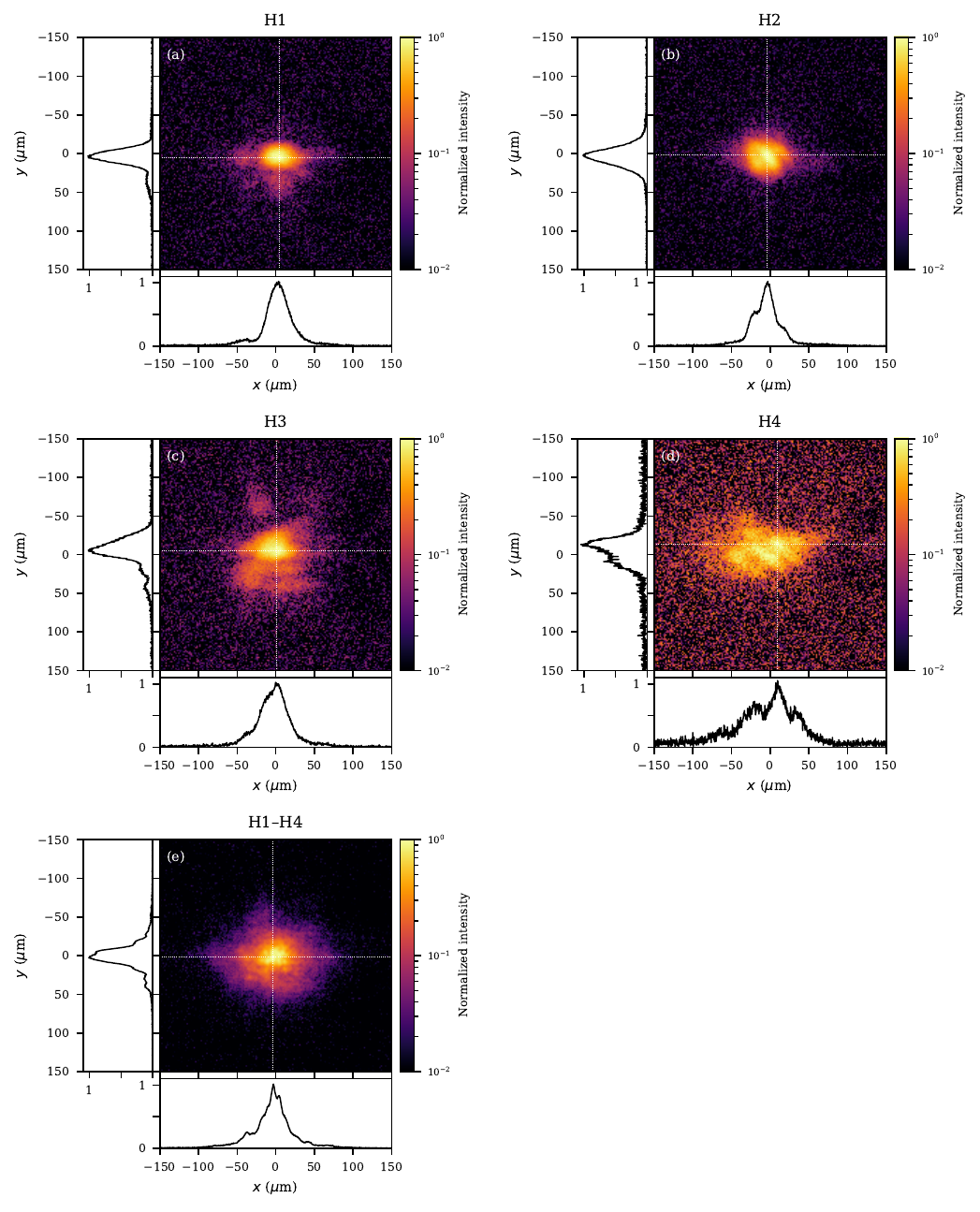}
  \caption{\rev{
  \textbf{Far-field images of the alignment laser at the simulated target position.}
  Background-subtracted images recorded at the position that simulates the target position for beams H1--H4 (a--d) and for the four beams together (e), all with the same exposure.
  Each panel is shown on a logarithmic color scale normalized to its own peak (the maximum after averaging over $5\times5$ pixels).
  The curves to the left of and below each image are the profiles through the peak (dotted lines) along $y$ and $x$, averaged over 5 pixels across the line, on the same normalization with a linear scale.
  The origin is the position of the sphere, determined from its shadow in the images recorded with the sphere in place, and the pixel scale of $0.34~\mu$m was calibrated with the shadow of the 105.6-$\mu$m sphere (uncertainty 10\%).
  }}
  \label{fig:ffp}
\end{figure*}

\rev{The sphere was positioned at the target position to within $5~\mu$m with the target-alignment system.}
\rev{Figure~\ref{fig:ffp} shows the far-field images of the alignment laser at the simulated target position for each beam and for the four beams together.}

Shot-to-shot pointing fluctuations displace the focal spot with respect to the sphere and could also affect the measurement.
\rev{These fluctuations stay within $\pm10~\mu$rad on target, i.e., $\pm40~\mu$m at the focal plane.}
A displacement of the beam centroid reduces the energy blocked by the sphere and therefore biases the inferred spot size toward larger values; the reported $D_{50}$ values thus include the contribution of residual pointing errors.
\rev{The displacement of each shot was also included in the fit as a normal random variable, with the recorded range taken as two standard deviations ($20~\mu$m per axis).}
\rev{The resulting single-shot $D_{50}$ values are less than 10\% smaller, and the peak intensities less than 20\% higher, than the reported values.}
Because target experiments are subject to the same pointing statistics, the values reported here represent the spot size effectively delivered to a target.

\begin{figure*}[!tbp]
  \raggedright
  \hspace{0.06\linewidth}
  \textbf{\fontsize{14pt}{14pt}\selectfont H1} \hspace{0.445\linewidth} \textbf{\fontsize{14pt}{14pt}\selectfont H2} \\ \vspace{5pt}

  \begin{minipage}[h]{0.49\linewidth}
  \centering
  \includegraphics[width=0.8\linewidth]{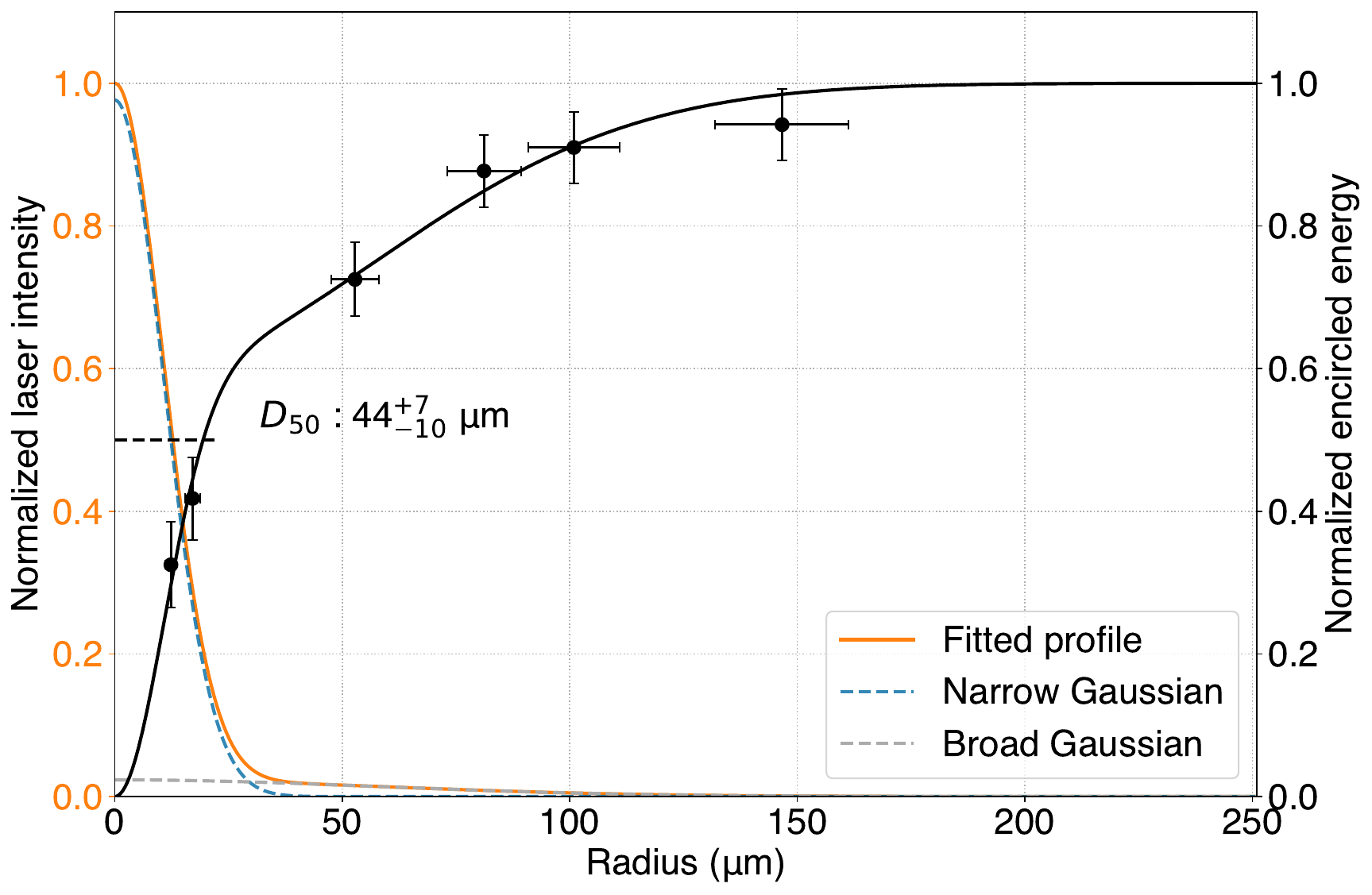}
  \end{minipage}
  \begin{minipage}[h]{0.49\linewidth}
  \centering
  \includegraphics[width=0.8\linewidth]{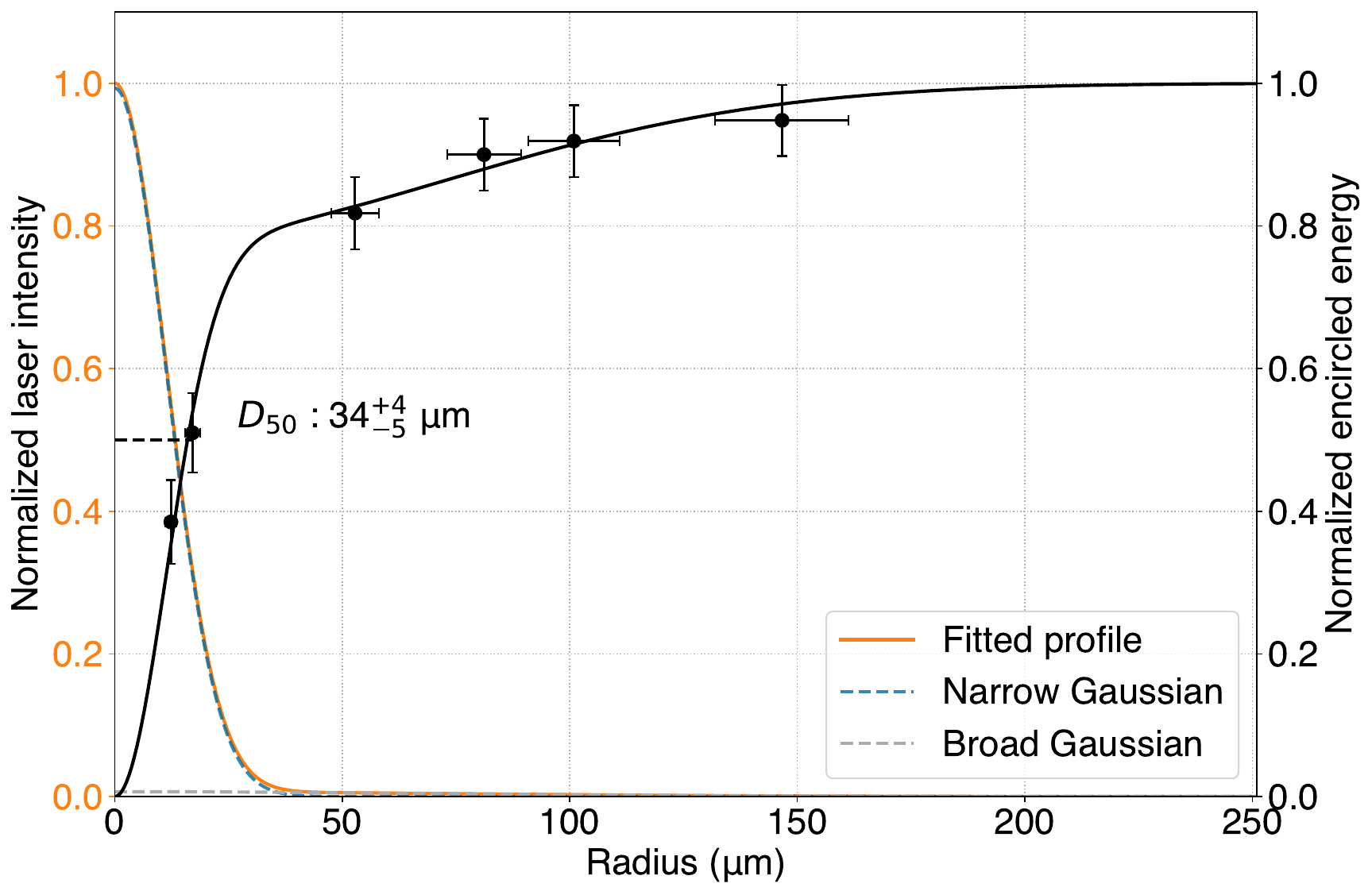}
  \end{minipage}
  \\ \vspace{15pt}

  \raggedright
  \hspace{0.06\linewidth}
  \textbf{\fontsize{14pt}{14pt}\selectfont H3} \hspace{0.445\linewidth} \textbf{\fontsize{14pt}{14pt}\selectfont H4} \\ \vspace{5pt}

  \begin{minipage}[h]{0.49\linewidth}
  \centering
  \includegraphics[width=0.8\linewidth]{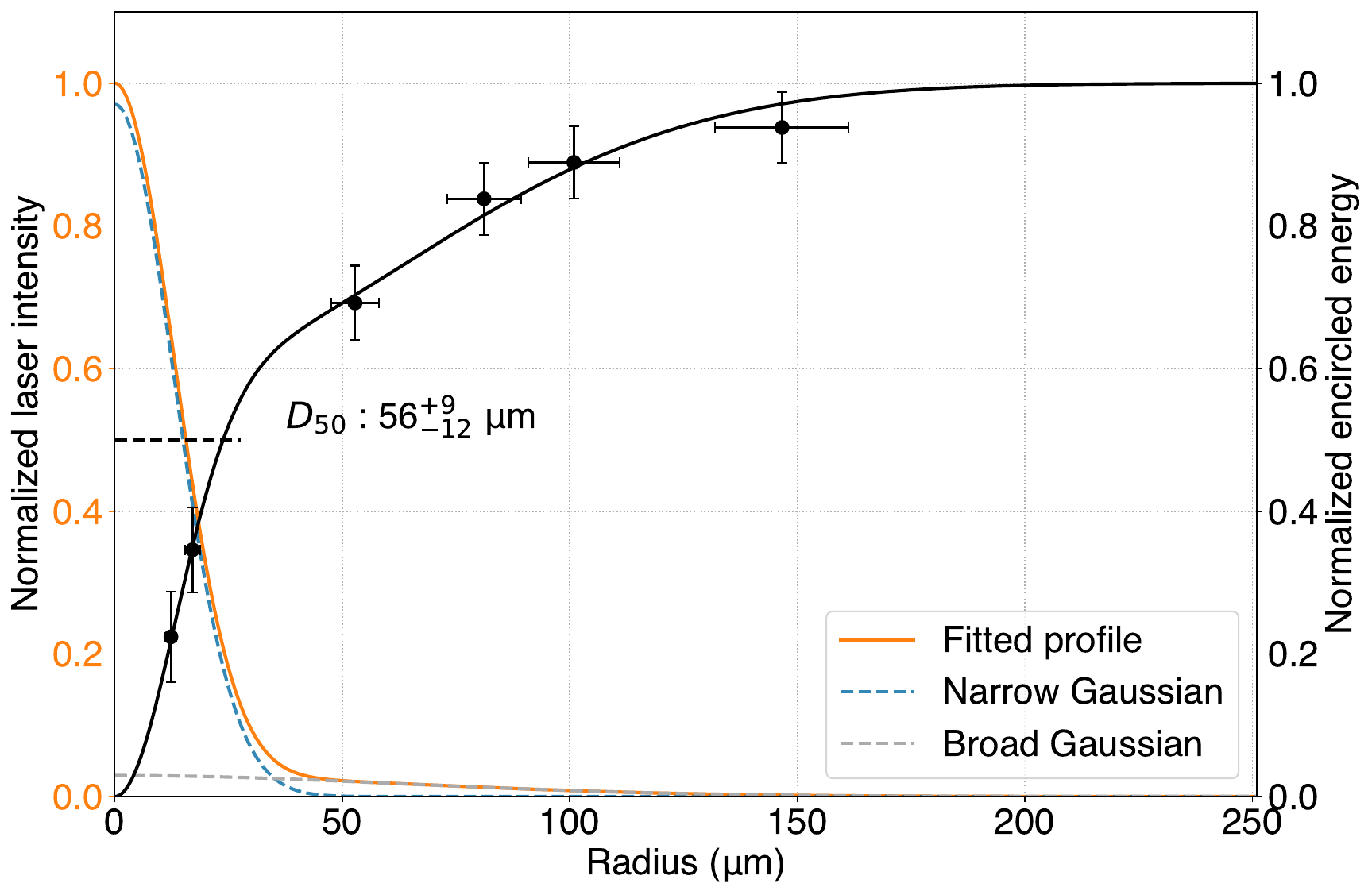}
  \end{minipage}
  \begin{minipage}[h]{0.49\linewidth}
  \centering
  \includegraphics[width=0.8\linewidth]{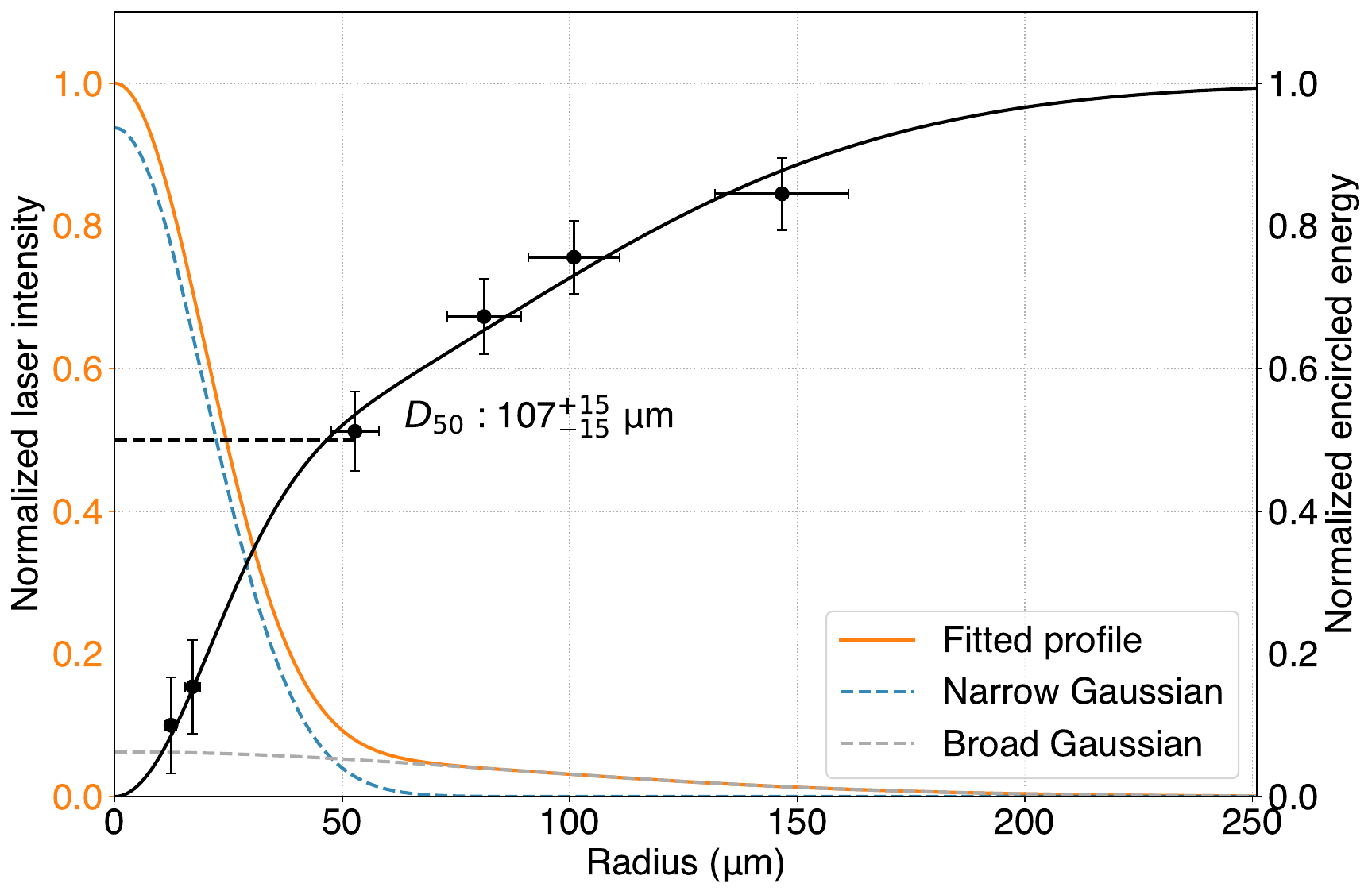}
  \end{minipage}

  \caption{
  \textbf{Focal spot of the alignment laser.}
    With all deformable mirrors (DFMs) corrected in closed loop, the alignment laser \rev{is focused to $D_{50} = 34$--$56~\mu$m for H1--H3, three to six times the diffraction-limited $D_{50}$ of $10~\mu$m.}
    The horizontal axis is \rev{the radial coordinate}, and the vertical axis shows the normalized intensity and normalized encircled energy.
    \rev{Dots indicate the measurements, with horizontal error bars showing the 10\% uncertainty of the sphere radius and vertical error bars the transmittance uncertainty; the orange and black curves show the fitted laser intensity and encircled energy at the MAP parameters, and the quoted $D_{50}$ is the posterior median with its 68\% HPD interval.}
    The dotted lines represent the ``narrow'' and ``broad'' components of the laser profile.
  }
  \label{fig:AL}
\end{figure*}

We inferred the spatial distribution of the laser fluence from the measured transmittance using Bayesian inference.
Note that the \rev{calorimeter} records time-integrated energy; the quantity reconstructed in this measurement is therefore the spatial distribution of the laser fluence at the focal plane, from which the spot size is estimated.
Assuming that the spatial profile of the beam does not vary during the pulse, the normalized radial profile of the fluence is identical to that of the intensity, and we denote it by \(I(r)\).
\rev{The encircled energy of the alignment laser rises steeply at small radii and then approaches unity only slowly (Fig.~\ref{fig:AL}), a shape that a single Gaussian cannot reproduce.}
The profile \(I(r)\) was modeled as a two-component Gaussian mixture comprising ``narrow'' and ``broad'' components:
\begin{align}
I(r)=\qty(1-S)\exp(-\frac{2r^{2}}{w_{n}^{2}})
+S\exp(-\frac{2r^{2}}{w_{b}^{2}}),
\end{align}
where \(w_n\) and \(w_b\) are the \(1/e^{2}\) radii of the respective components, and \(S\) is the mixing fraction of the broad component \((0\le S\le 1)\).
The corresponding normalized encircled energy, \(E_{\mathrm{enc}}(r)\), is
\begin{align}
E_{\mathrm{enc}}(r)=
\frac{\qty(1-S)w_{n}^{2}\qty(1 - e^{-2r^{2}/w_{n}^{2}})+Sw_{b}^{2}\qty(1 - e^{-2r^{2}/w_{b}^{2}})}
{\qty(1-S)w_{n}^{2}+Sw_{b}^{2}},
\end{align}
which we compared directly with the experimental data.
Furthermore, the energy fractions for each component are
\begin{align}
f_{\mathrm{narrow}}&=\frac{(1-S)\,w_{n}^{2}}{(1-S)\,w_{n}^{2}+S\,w_{b}^{2}},\qquad \\
f_{\mathrm{broad}}&=\frac{S\,w_{b}^{2}}{(1-S)\,w_{n}^{2}+S\,w_{b}^{2}}.
\end{align}
Even when the broad component has a small on-axis intensity ($S\ll1$), it can still carry a sizable fraction of the total energy if $w_{b}\gg w_{n}$.

\rev{The two components describe the core of the spot and the halo around it, which arise from different spatial frequencies of the wavefront error.}
\rev{A ripple with a spatial period $\Lambda$ across the aperture diffracts light to a radius $r = \lambda f/\Lambda$ in the focal plane.}
\rev{The narrow component ($w_n = 20$--$26~\mu\mathrm{m}$ for the alignment laser) thus contains the well-corrected part of the beam together with low-order aberrations ($\Lambda > 150~\mathrm{mm}$).}
\rev{The broad component ($w_b = 100$--$180~\mu\mathrm{m}$) collects the light scattered by mid-spatial-frequency errors ($\Lambda = 25$--$40~\mathrm{mm}$).}

In contrast to least-squares fitting, which returns a single best-fit parameter set, Bayesian inference treats the model parameters as random variables and yields their posterior distribution, i.e., the probability distribution of the parameters conditioned on the measured data.
Uncertainties and correlations among the parameters are thereby propagated to derived quantities without linear approximation, and asymmetric uncertainty intervals are obtained naturally.
This approach is well suited to the present data set, which consists of a limited number of shots and exhibits strong correlations among the model parameters (see the Supplementary Note).
Bayesian modeling was performed with \texttt{PyMC} (v5) using weakly informative uniform priors on \(S\) (\(0\le S\le 1\)), \(w_n\), and \(w_b\), with the ordering constraint \(w_n<w_b\) imposed so that the two components are identifiable.
Posterior summaries and 68\% highest posterior density (HPD) intervals, i.e., the narrowest intervals containing 68\% of the posterior probability, were computed with \texttt{ArviZ}.
\rev{The tables list posterior medians with 68\% HPD intervals, with derived quantities such as $D_{50}$, the energy fractions, and the peak intensity computed from each posterior sample.}
\rev{The curves in the figures are drawn at the maximum a posteriori (MAP) parameters.}

The beam diameter was defined from the half-encircled-energy radius \(r_{1/2}\) satisfying \(E_{\mathrm{enc}}(r_{1/2})=0.5\):
\begin{align}
D_{50} \equiv 2r_{1/2}.
\end{align}
\rev{The peak intensity for a pulse of energy $E_L$ and duration $\tau$ follows from the normalized profile as}
{\revcolor
\begin{align}
I_{\mathrm{peak}} &= \frac{E_L}{\tau A_{\mathrm{eff}}}, \nonumber\\
A_{\mathrm{eff}} &= \int_0^\infty I(r)\,2\pi r\,dr = \frac{\pi}{2}\qty[(1-S)w_n^{2}+S w_b^{2}],
\end{align}}
\rev{where $A_{\mathrm{eff}}$ is the effective area of the profile and $\tau$ is the pulse duration (FWHM).}
\rev{The peak intensity is quoted for $E_L = 250$~J per beam and $\tau = 1.3$~ps, which are typical of LFEX operation, and scales linearly with $E_L/\tau$.}

\begin{table*}[!tbp]
\centering
\begin{tabular}{lcccc}
\toprule
Parameter & H1 & H2 & H3 & H4\\
\midrule
$S$ & \rev{$0.030^{+0.013}_{-0.023}$} & \rev{$0.017^{+0.009}_{-0.016}$} & \rev{$0.043^{+0.021}_{-0.034}$} & \rev{$0.072^{+0.037}_{-0.061}$} \\
$w_n$ ($\mu$m) & \rev{$20^{+6}_{-5}$} & \rev{$21^{+5}_{-4}$} & \rev{$26^{+7}_{-7}$} & \rev{$54^{+17}_{-18}$} \\
$w_b$ ($\mu$m) & \rev{$102^{+19}_{-26}$} & \rev{$105^{+29}_{-41}$} & \rev{$113^{+23}_{-29}$} & \rev{$177^{+37}_{-58}$} \\
$D_{50}$ ($\mu$m) & \rev{$44^{+7}_{-10}$} & \rev{$34^{+4}_{-5}$} & \rev{$56^{+9}_{-12}$} & \rev{$107^{+15}_{-15}$} \\
$f_{\mathrm{narrow}}$ (\%) & \rev{$53^{+15}_{-16}$} & \rev{$68^{+18}_{-12}$} & \rev{$53^{+18}_{-14}$} & \rev{$51^{+20}_{-14}$} \\
$f_{\mathrm{broad}}$ (\%) & \rev{$47^{+16}_{-15}$} & \rev{$32^{+12}_{-18}$} & \rev{$47^{+14}_{-18}$} & \rev{$49^{+14}_{-20}$} \\
\rev{$I_{\mathrm{peak}}$ ($10^{18}$ W/cm$^2$)} & \rev{$16^{+4}_{-6}$} & \rev{$20^{+4}_{-6}$} & \rev{$9.8^{+2.6}_{-3.9}$} & \rev{$2.3^{+0.7}_{-1.1}$} \\
\bottomrule
\end{tabular}
\caption{Fitted parameters for each beam of the alignment laser \rev{(posterior medians with 68\% HPD intervals). The peak intensity $I_{\mathrm{peak}}$ is the value that the profile would give for $E_L = 250$~J and $\tau = 1.3$~ps, i.e., the intensity attainable if the full-power beam had the focusing quality of the alignment laser.}}
\label{tab:AL}
\end{table*}

Figure~\ref{fig:AL} shows the focal spot of the alignment laser for each beam, i.e., the focusing performance without \rev{the thermal wavefront distortion of the main amplifier}.
The corresponding parameters are summarized in Table~\ref{tab:AL}.
\rev{The narrow component of H1, H2, and H3 has a waist of 20--26~$\mu$m, and the smallest spot, $D_{50} = 34^{+4}_{-5}~\mu\mathrm{m}$ for H2, is 3.4 times the diffraction-limited $D_{50}$ of $10~\mu\mathrm{m}$.}
The beam aperture of the H4 beam was reduced to mask the optical damage, resulting in a larger spot size than for the other beams.
\rev{The far-field images (Fig.~\ref{fig:ffp}) show a compact core with a weak halo for H1 and H2, a core surrounded by a fringed halo for H3, and a broad fringed pattern without a distinct core for H4.}

\rev{The on-axis intensity fraction of the broad component, $S$, is only 3--4\% for H1 and H3, 2\% for H2, and 7\% for H4.}
\rev{Nevertheless, the broad component carries 47\% (H1 and H3), $32^{+12}_{-18}$\% (H2), and $49^{+14}_{-20}$\% (H4) of the total energy, because $w_b$ is three to five times $w_n$.}
\rev{The smaller broad-component fraction of H2 is attributed to the deformable mirror upstream of the compressor, which removes the wavefront error of the compressor and the final optics.}
\rev{The difference between H2 and H1/H3, about 15 percentage points, is thus the fraction of the energy that these optics scatter into the broad component on the beams without this mirror.}
\rev{The peak intensities that the full-power beams would reach with the focusing quality of the alignment laser are $1$--$2\times10^{19}~\mathrm{W/cm^2}$ for H1--H3 at 250~J and 1.3~ps (Table~\ref{tab:AL}).}

\begin{figure*}[!tbp]
  \raggedright
  \hspace{0.06\linewidth}
  \textbf{\fontsize{14pt}{14pt}\selectfont H1} \hspace{0.41\linewidth} \textbf{\fontsize{14pt}{14pt}\selectfont H2} \\ \vspace{5pt}

  \begin{minipage}[h]{0.49\linewidth}
  \centering
  \includegraphics[width=0.8\linewidth]{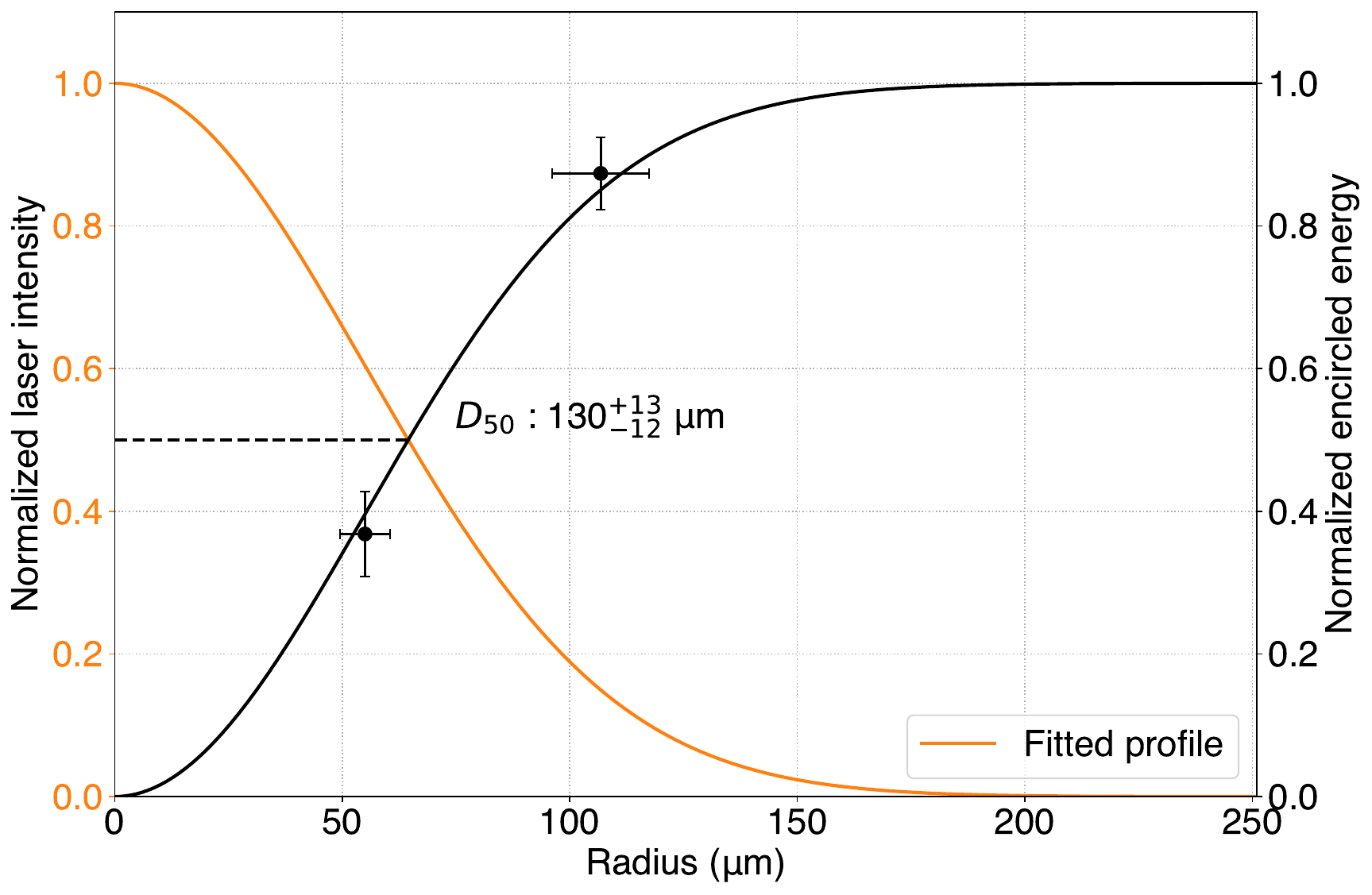}
  \end{minipage}
  \begin{minipage}[h]{0.49\linewidth}
  \centering
  \includegraphics[width=0.8\linewidth]{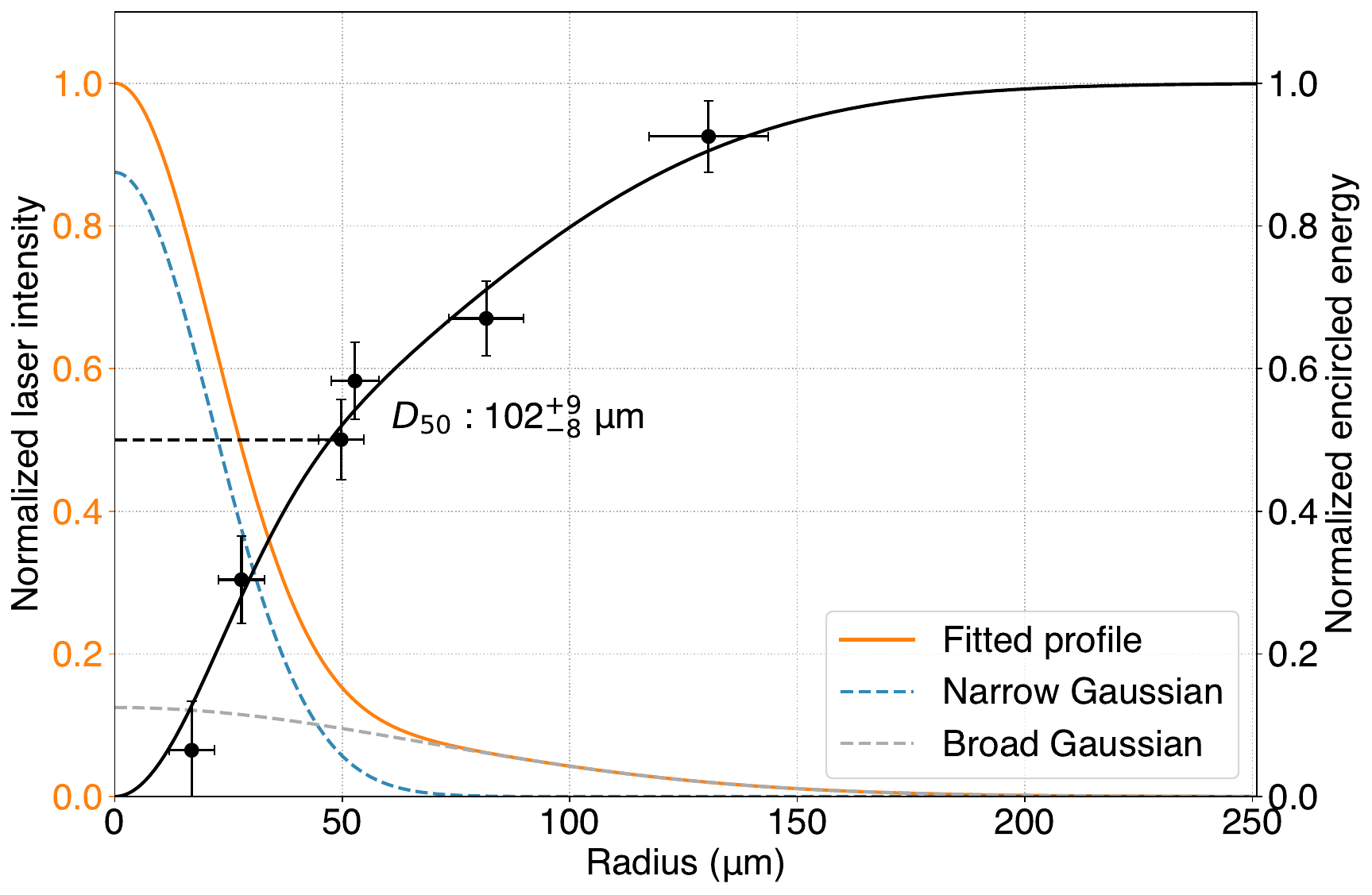}
  \end{minipage}
  \\ \vspace{15pt}

  \raggedright
  \hspace{0.06\linewidth}
  \textbf{\fontsize{14pt}{14pt}\selectfont H3} \hspace{0.41\linewidth} \textbf{\fontsize{14pt}{14pt}\selectfont H4} \\ \vspace{5pt}

  \begin{minipage}[h]{0.49\linewidth}
  \centering
  \includegraphics[width=0.8\linewidth]{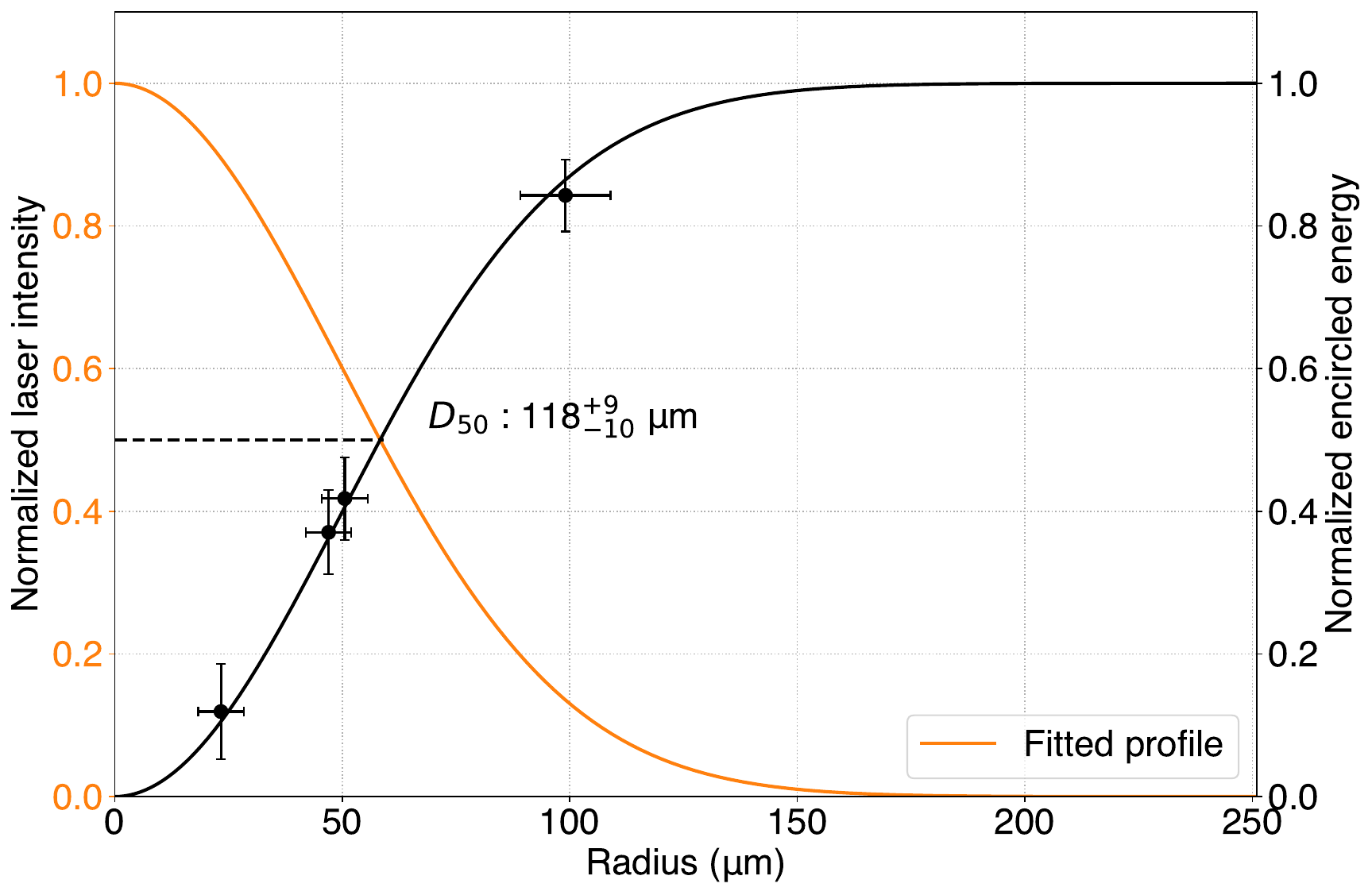}
  \end{minipage}
  \begin{minipage}[h]{0.49\linewidth}
  \centering
  \includegraphics[width=0.8\linewidth]{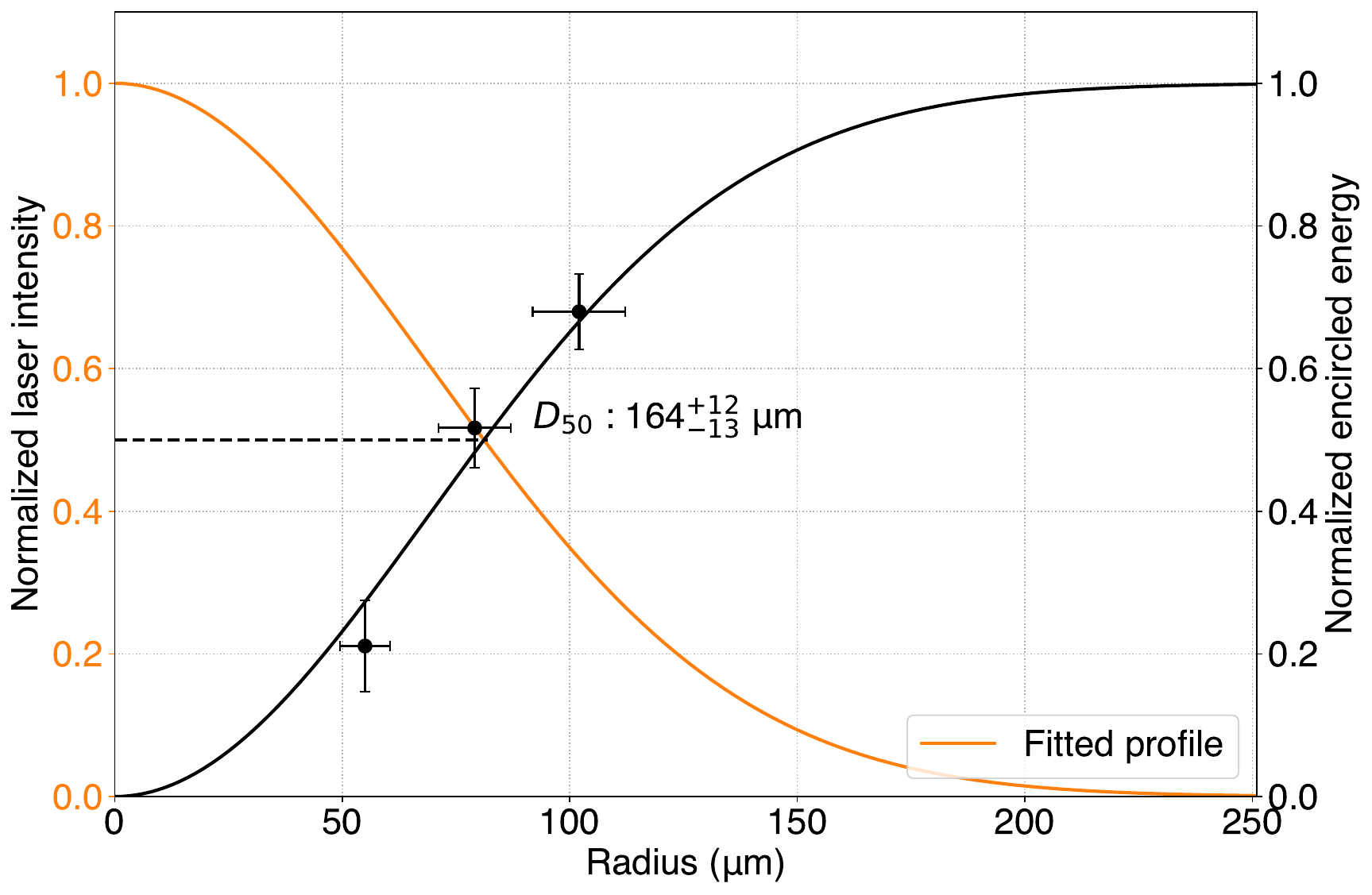}
  \end{minipage}

  \caption{
  \textbf{Focal spot for full-power operation.}
  Laser-spot measurements taken with thermal distortion in the main amplifier set to a level equivalent to full-power operation.
  The horizontal axis is \rev{the radial coordinate}, and the vertical axis shows the normalized intensity and normalized encircled energy.
  \rev{Dots indicate the measurements, with horizontal error bars showing the uncertainty of the effective blocking radius, $\max(5~\mu\mathrm{m},\,0.1\,r_\mathrm{sphere})$, and vertical error bars the transmittance uncertainty; the orange and black curves show the fitted laser intensity and encircled energy at the MAP parameters, and the quoted $D_{50}$ is the posterior median with its 68\% HPD interval.}
  The dotted lines represent the ``narrow'' and ``broad'' components of the laser profile.
  For beams other than H2, the spot was larger and \rev{the two components could not be separated}; the fitted curves for H1, H3, and H4 are therefore single-Gaussian fits.
  }
  \label{fig:DS}
\end{figure*}

\begin{table}[!htbp]
\centering
\setlength{\tabcolsep}{4pt}
\begin{tabular}{lcccc}
\toprule
Parameter & H1 & H2 & H3 & H4 \\
\midrule
$S$ & $1.0$ & \rev{$0.19^{+0.12}_{-0.19}$} & $1.0$ & $1.0$ \\
$w_n\;(\mu\mathrm{m})$ & $-$ & \rev{$51^{+22}_{-20}$} & $-$ & $-$ \\
$w_b\;(\mu\mathrm{m})$ & \rev{$110^{+11}_{-10}$} & \rev{$120^{+20}_{-34}$} & \rev{$100^{+7}_{-8}$} & \rev{$139^{+11}_{-11}$} \\
$D_{50}$ ($\mu\mathrm{m}$) & \rev{$130^{+13}_{-12}$} & \rev{$102^{+9}_{-8}$} & \rev{$118^{+9}_{-10}$} & \rev{$164^{+12}_{-13}$} \\
$f_{\mathrm{narrow}}$ (\%) & $-$ & \rev{$38^{+12}_{-38}$} & $-$ & $-$ \\
$f_{\mathrm{broad}}$ (\%) & $-$ & \rev{$62^{+38}_{-12}$} & $-$ & $-$ \\
\rev{$I_{\mathrm{peak}}$ ($10^{18}$ W/cm$^2$)} & \rev{$1.0^{+0.2}_{-0.2}$} & \rev{$2.2^{+0.6}_{-0.9}$} & \rev{$1.2^{+0.2}_{-0.2}$} & \rev{$0.63^{+0.09}_{-0.11}$} \\
\bottomrule
\end{tabular}
\caption{Fitted parameters for each beam under the full-power condition \rev{(posterior medians with 68\% HPD intervals)}. For H1, H3, and H4, the two-component \rev{model is not identifiable}, and the quoted values are those of a single-Gaussian fit (equivalent to $S=1$), in which case $w_b$ denotes the $1/e^{2}$ radius of the single Gaussian. \rev{The peak intensity $I_{\mathrm{peak}}$ is given for $E_L = 250$~J and $\tau = 1.3$~ps.}}
\label{tab:DS}
\end{table}

Figure~\ref{fig:DS} shows the laser focal spot obtained under the full-power shot condition.
The corresponding parameters are summarized in Table~\ref{tab:DS}.
\rev{The spot diameters were large, and for the H1, H3, and H4 beams the two-component model is not identifiable (see the Supplementary Note).}
For these beams we therefore adopted a single-Gaussian fit, which corresponds to the two-component model with $S=1$.

\rev{The H2 beam, whose deformable mirror is installed before the compressor, retains a narrow component.}
\rev{Its width $w_n = 51^{+22}_{-20}~\mu\mathrm{m}$ is, however, 2.5 times that of the alignment laser, and the energy fraction of the broad component increased from 32\% to $62^{+38}_{-12}$\%.}
\rev{Both changes are attributed to the thermal wavefront distortion of the main amplifier, which the deformable mirrors inside the amplifier do not fully compensate.}

\rev{The full-power $D_{50}$ values are 10--16 times the diffraction-limited value of the full aperture, compared with 3.4--11 times for the alignment laser.}
\rev{The $D_{50}$ values of H1/H3 and H2 are similar, but the narrow component of H2 raises its peak intensity to $2.2^{+0.7}_{-1.0}$ and $1.8^{+0.5}_{-0.8}$ times those of H1 and H3, respectively (Table~\ref{tab:DS}).}
\rev{The full-power condition reduces the peak intensity relative to the alignment laser by a factor of $16^{+5}_{-8}$ (H1), $9^{+4}_{-5}$ (H2), $8^{+3}_{-3}$ (H3), and $3.6^{+1.2}_{-1.9}$ (H4).}

%%%%%%%%%%%%%%%%%%%%%%%%%%%%%%%%%%%%%%%%%%%%%%%%%%%%%%%%%%%%%%%%%%%%%%%%%%%%%%%%%%%%%%%%%

\section{\texorpdfstring{\rev{Scope of the method}}{Scope of the method}}
\label{sec:discussion}

\rev{The present method differs in kind from the focal-spot diagnostics of other kilojoule-class petawatt facilities.}
\rev{Those diagnostics reconstruct the on-shot spot from wavefront measurements of a sampled beam (OMEGA-EP\cite{Bromage2008-as,Kruschwitz2012-ls}, NIF-ARC\cite{Williams2021-ls}, PHELIX\cite{Ohland2022-ls}) or image the harmonic emission of the target surface (PETAL, Vulcan\cite{Raffestin2019-pt,Dromey2009-ls}).}
\rev{They yield the two-dimensional spot.}
\rev{The wavefront measurements also give the aberrations that produce it but require sampling optics at the final optical stage.}
\rev{Blocking the core of the spot with an opaque ball to record the halo is an established technique in the focal-spot imaging of high-energy lasers\cite{Wang2017-ls}; the present method adapts it to a destructive single-shot beam by scanning the sphere diameter.}
\rev{It yields only the azimuthally averaged encircled energy of each beam, and its resolution at small radii is limited by the smallest sphere ($25~\mu$m in diameter) and by the plasma expansion (about $5~\mu$m).}
\rev{In exchange, it needs no sampling optics or wavefront sensor at the final optical stage, which LFEX does not have, and it measures the energy that reaches the target plane through the full-aperture focusing optic.}
\rev{The method is therefore suited to providing facility users with a reference value of the spot size and of the on-target intensity, rather than to diagnosing aberrations.}

\rev{An upgrade of LFEX now in progress will add a focal-spot diagnostic similar to that of OMEGA-EP, based on the wavefront of the beam sampled after compression.}
\rev{The spots measured with this diagnostic will be compared with the present measurement.}
\rev{The combined four-beam spot, the quantity most relevant to experiments, will also be measured then.}
\rev{This spot is further broadened by the pointing differences between the beams (up to $20~\mu$rad, or $80~\mu$m at the focal plane).}

%%%%%%%%%%%%%%%%%%%%%%%%%%%%%%%%%%%%%%%%%%%%%%%%%%%%%%%%%%%%%%%%%%%%%%%%%%%%%%%%%%%%%%%%%

\section{Conclusion}

\rev{Encircled-energy distributions were measured with spherical occluders under full-power conditions and for the alignment laser, and the focal-spot parameters were inferred by Bayesian inference.}
\rev{For the alignment laser, which is free of the thermal distortion of the main amplifier, the narrow component carries 53\% of the energy for H1 and H3 and 68\% for H2.}
\rev{The smallest $D_{50}$ ($34~\mu$m, H2) is 3.4 times the diffraction-limited value of $10~\mu$m.}
The H2 beam line includes a deformable mirror (DFM400) upstream of the pulse compressor, which enhances the focusing performance relative to the other beams; H4 exhibits poorer focusing because its aperture is intentionally restricted to mask optical damage.
\rev{On full-power shots, the thermal distortion of the main amplifier broadens $D_{50}$ to 102--164~$\mu$m, 10--16 times the diffraction limit of the full aperture.}
\rev{The peak intensities for 250~J and 1.3~ps are $1.0$, $2.2$, $1.2$, and $0.63\times10^{18}~\mathrm{W/cm^{2}}$ for H1--H4.}
\rev{These intensities are a factor of 8--16 (H1--H3) and 3.6 (H4) below the intensities that the same beams would reach with the focusing quality of the alignment laser.}
\rev{These values include the shot-to-shot pointing fluctuations and represent the spot delivered to a target over a series of shots.}
\rev{The DFM400 of H2 doubles the on-shot peak intensity relative to H1 and H3 at a similar $D_{50}$;} future upgrades will equip all beams with DFM400 deformable mirrors.

\section*{Acknowledgments}
The authors thank the technical support staff at The University of Osaka for assistance with the laser operation, target fabrication, plasma diagnostics, and computer simulation.
This work was partially achieved through the use of large-scale computer systems at the Cybermedia Center at The University of Osaka.
This work is supported by the Joint Research Project/Collaboration Research Program between the National Institute for Fusion Science and the Institute of Laser Engineering at The University of Osaka, a Grant-in-Aid for Scientific Research (No.~25K17369), %Takizawa
the JSPS Core-to-Core Program (Grant No.~JPJSCCA20230003), and ``Power Laser DX Platform'' as research equipment shared in MEXT Project for promoting public utilization of advanced research infrastructure (Program for advanced research equipment platforms) Grant Number JPMXS0450300021.

\section*{DATA AVAILABILITY}

The data that support the findings of this study are available from the corresponding author upon reasonable request.

\section*{Supplementary Note: Posterior distributions}

% Supplementary figure numbering: S1, S2, ...
\renewcommand{\thefigure}{S\arabic{figure}}
\renewcommand{\theHfigure}{S\arabic{figure}}
\setcounter{figure}{0}

To assess parameter identifiability and correlations in the two-component model, we provide joint posterior distributions for $(S,w_{n},w_{b})$ of all four LFEX beams under the full-power condition (Supplementary Fig.~S1).
\rev{For H1, H3, and H4 the two-component model is not identifiable: the sampler converges ($\hat{R} = 1.00$, no divergent transitions), but the posterior of $S$ extends over almost the whole prior range and $w_n$ is unconstrained (panels a, c, and d).}
\rev{For these beams the values quoted in Table~\ref{tab:DS} are those of the single-Gaussian fit.}

\begin{figure*}[!tbp]
  \centering

  % Row 1
  \begin{minipage}[b]{0.485\linewidth}
    \centering
    \includegraphics[width=\linewidth]{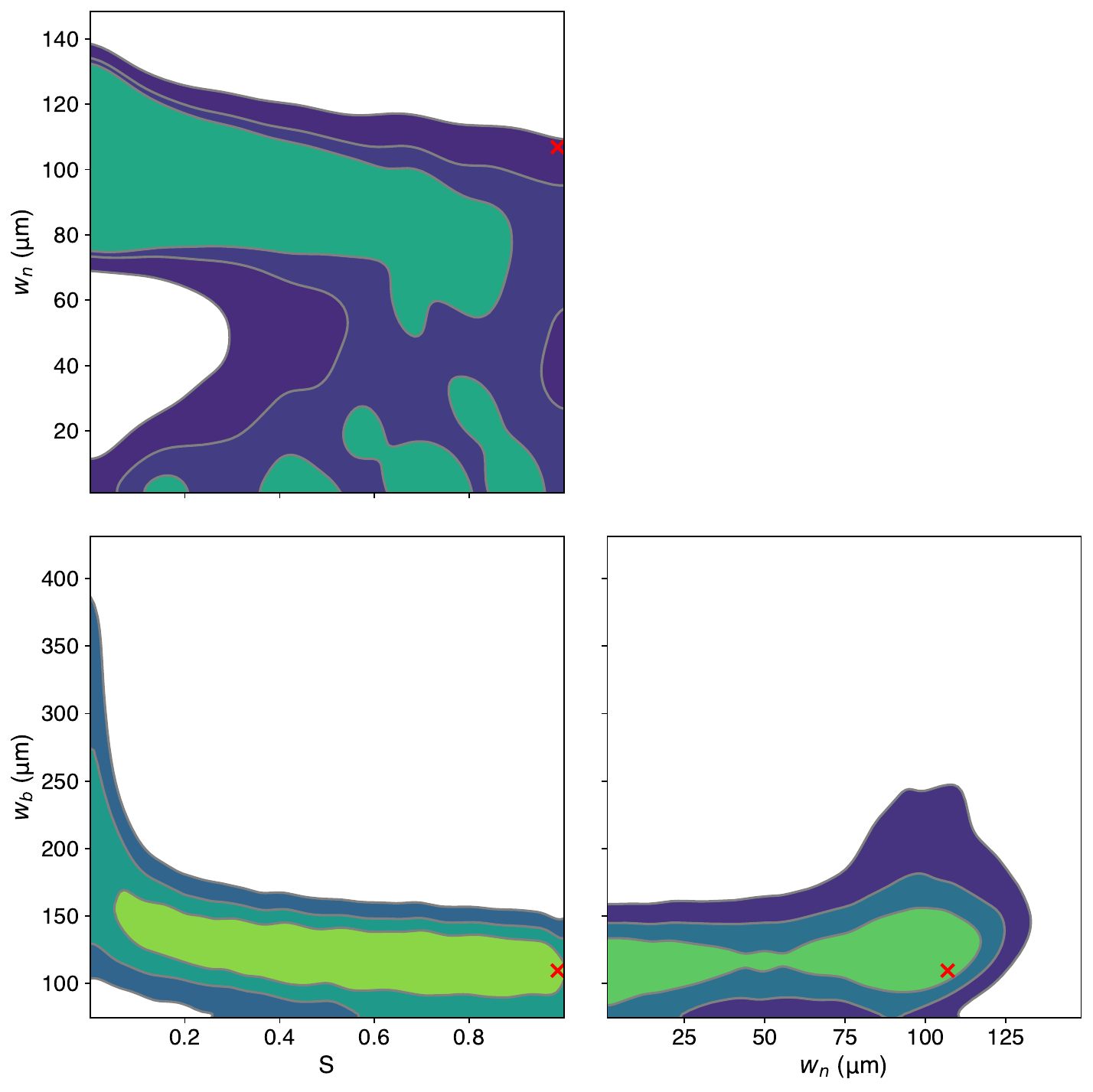}
    \vspace{0.4ex}
    \small\textbf{(a) H1}
  \end{minipage}\hfill
  \begin{minipage}[b]{0.485\linewidth}
    \centering
    \includegraphics[width=\linewidth]{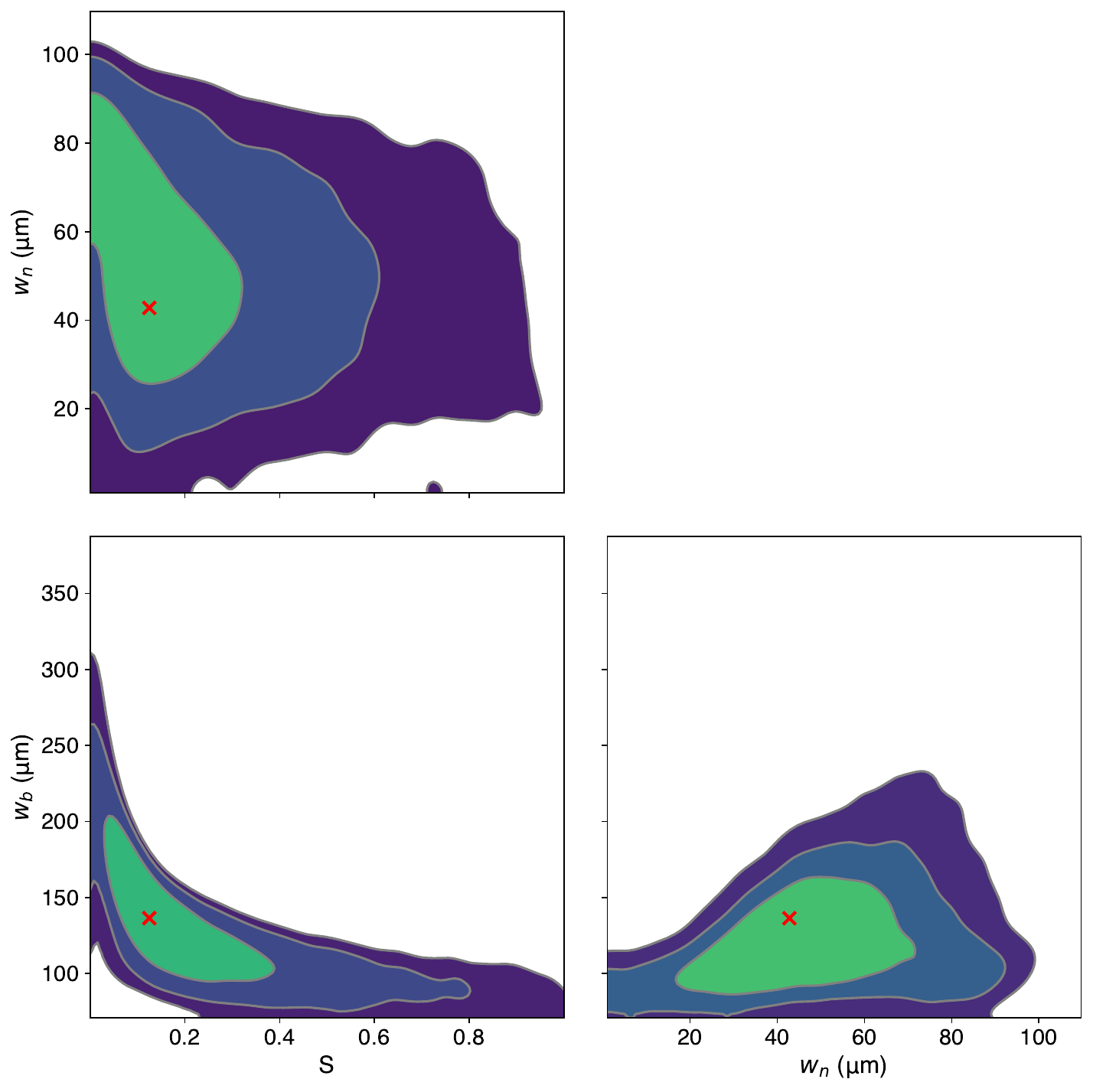}
    \vspace{0.4ex}
    \small\textbf{(b) H2}
  \end{minipage}

  \vspace{0.9ex}

  % Row 2
  \begin{minipage}[b]{0.485\linewidth}
    \centering
    \includegraphics[width=\linewidth]{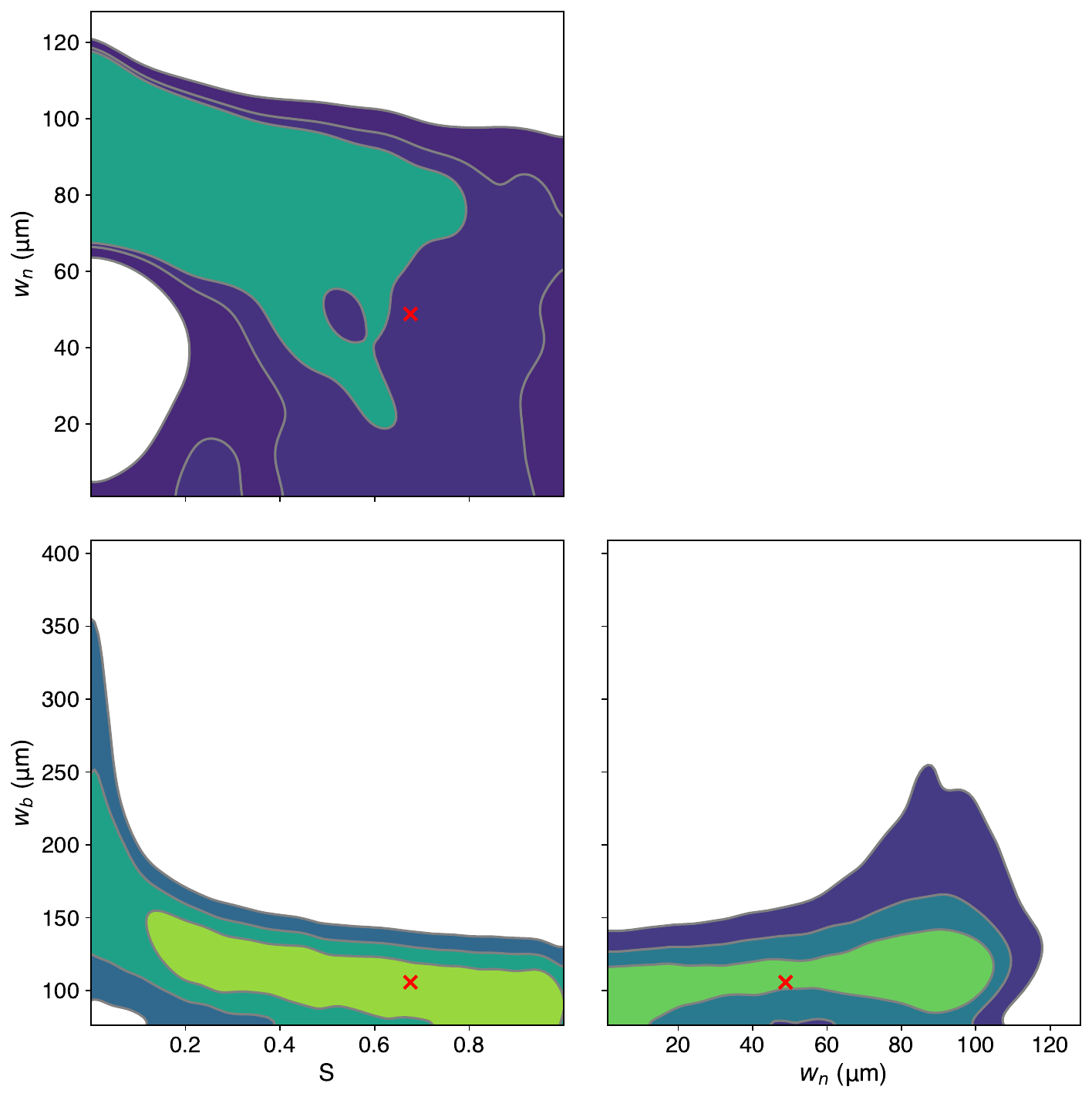}
    \vspace{0.4ex}
    \small\textbf{(c) H3}
  \end{minipage}\hfill
  \begin{minipage}[b]{0.485\linewidth}
    \centering
    \includegraphics[width=\linewidth]{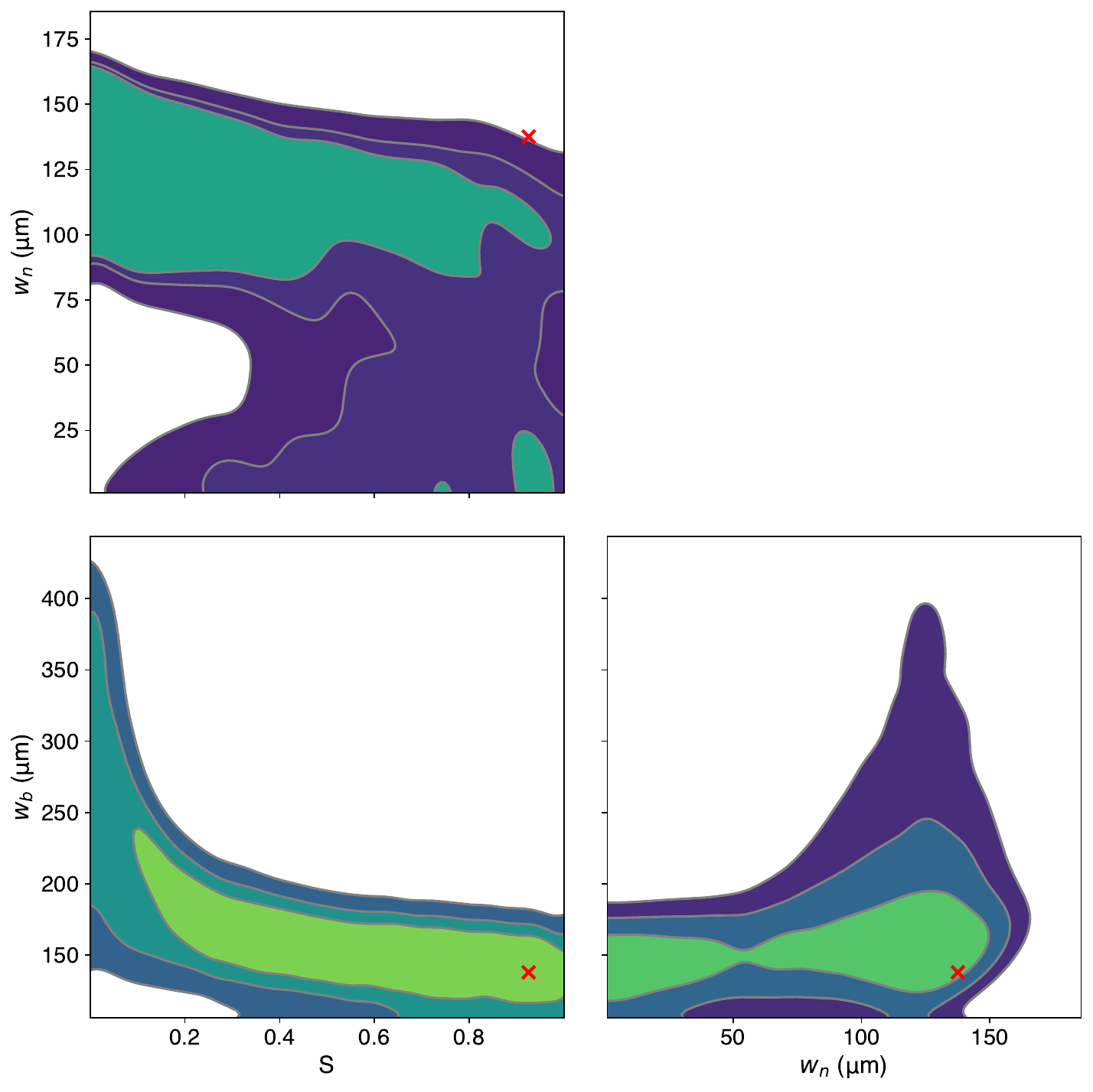}
    \vspace{0.4ex}
    \small\textbf{(d) H4}
  \end{minipage}

  \caption{\textbf{Joint posterior distributions for all beams (H1--H4) under full-power conditions.}
  Each panel shows the two-dimensional marginals for $(S,w_n)$, $(S,w_b)$, and $(w_n,w_b)$ derived from the Bayesian fit in the main text.
  Equal-density contours indicate higher posterior probability; the red cross marks the MAP estimate of the two-component model.
  For H1, H3, and H4, where the two-component \rev{model is not identifiable}, these crosses do not correspond to the values in Table~\ref{tab:DS}, which are from single-Gaussian fits.
  }
  \label{fig:supp_post_all}
\end{figure*}

\FloatBarrier % Keep the supplementary figure before the references.

\bibliographystyle{hplse}
\bibliography{ref}

\begin{thebibliography}{10}
\newcommand{\enquote}[1]{``#1''}
\expandafter\ifx\csname urlstyle\endcsname\relax
  \providecommand{\doi}[1]{doi:\discretionary{}{}{}#1}\else
  \providecommand{\doi}{doi:\discretionary{}{}{}\begingroup
  \urlstyle{rm}\Url}\fi

\bibitem{Strickland1985-bo}
D.~Strickland and G.~Mourou, \enquote{Compression of amplified chirped optical
  pulses}, Opt. Commun. 56, 219--221 (1985). DOI:
  https://doi.org/10.1016/0030-4018(85)90120-8

\bibitem{Perry1999-ot}
M.~D. Perry, D.~Pennington, B.~C. Stuart, G.~Tietbohl, J.~A. Britten, C.~Brown,
  S.~Herman, B.~Golick, M.~Kartz, J.~Miller, H.~T. Powell, M.~Vergino, and
  V.~Yanovsky, \enquote{Petawatt laser pulses}, Opt. Lett. 24, 160--162 (1999).
  DOI: https://doi.org/10.1364/OL.24.000160

\bibitem{Danson2019-fg}
C.~N. Danson, C.~Haefner, J.~Bromage, T.~Butcher, J.-C.~F. Chanteloup, E.~A.
  Chowdhury, A.~Galvanauskas, L.~A. Gizzi, J.~Hein, D.~I. Hillier, N.~W. Hopps,
  Y.~Kato, E.~A. Khazanov, R.~Kodama, G.~Korn, R.~Li, Y.~Li, J.~Limpert, J.~Ma,
  C.~H. Nam, D.~Neely, D.~Papadopoulos, R.~R. Penman, L.~Qian, J.~J. Rocca,
  A.~A. Shaykin, C.~W. Siders, C.~Spindloe, S.~Szatmári, R.~M. G.~M. Trines,
  J.~Zhu, P.~Zhu, and J.~D. Zuegel, \enquote{Petawatt and exawatt class lasers
  worldwide}, High Power Laser Sci. Eng. 7, e54 (2019). DOI:
  https://doi.org/10.1017/hpl.2019.36

\bibitem{Miyanaga2006-bq}
N.~Miyanaga, H.~Azechi, K.~A. Tanaka, T.~Kanabe, T.~Jitsuno, J.~Kawanaka,
  Y.~Fujimoto, R.~Kodama, H.~Shiraga, K.~Knodo, K.~Tsubakimoto, H.~Habara,
  J.~Lu, G.~Xu, N.~Morio, S.~Matsuo, E.~Miyaji, Y.~Kawakami, Y.~Izawa, and
  K.~Mima, \enquote{10-{kJ} {PW} laser for the {FIREX}-{I} program}, J. Phys.
  IV 133, 81--87 (2006). DOI: https://doi.org/10.1051/jp4:2006133016

\bibitem{Waxer2008-sn}
L.~J. Waxer, M.~J. Guardalben, J.~H. Kelly, B.~E. Kruschwitz, J.~Qiao, I.~A.
  Begishev, J.~Bromage, C.~Dorrer, J.~L. Edwards, L.~Folnsbee, S.~D. Jacobs,
  R.~Jungquist, T.~J. Kessler, R.~W. Kidder, S.~J. Loucks, J.~R. Marciante,
  D.~N. Maywar, R.~L. McCrory, D.~D. Meyerhofer, S.~F.~B. Morse, A.~V. Okishev,
  J.~B. Oliver, G.~Pien, J.~Puth, A.~L. Rigatti, W.~Schmid, M.~J. Shoup,
  C.~Stoeckl, K.~A. Thorp, and J.~D. Zuegel, \enquote{The {OMEGA} {EP}
  high-energy, short-pulse laser system}, in \emph{Conference on Lasers and
  Electro-Optics/Quantum Electronics and Laser Science Conference and Photonic
  Applications Systems Technologies} (Optica Publishing Group, 2008), p. JThB1.

\bibitem{Barty2004-ib}
C.~P.~J. Barty, M.~Key, J.~Britten, R.~Beach, G.~Beer, C.~Brown, S.~Bryan,
  J.~Caird, T.~Carlson, J.~Crane, J.~Dawson, A.~C. Erlandson, D.~Fittinghoff,
  M.~Hermann, C.~Hoaglan, A.~Iyer, L.~Jones, I.~Jovanovic, A.~Komashko,
  O.~Landen, Z.~Liao, W.~Molander, S.~Mitchell, E.~Moses, N.~Nielsen, H.-H.
  Nguyen, J.~Nissen, S.~Payne, D.~Pennington, L.~Risinger, M.~Rushford,
  K.~Skulina, M.~Spaeth, B.~Stuart, G.~Tietbohl, and B.~Wattellier, \enquote{An
  overview of {LLNL} high-energy short-pulse technology for advanced
  radiography of laser fusion experiments}, Nucl. Fusion 44, S266 (2004). DOI:
  https://doi.org/10.1088/0029-5515/44/12/s18

\bibitem{Blanchot2017-xv}
N.~Blanchot, G.~Béhar, J.~C. Chapuis, C.~Chappuis, S.~Chardavoine, J.~F.
  Charrier, H.~Coïc, C.~Damiens-Dupont, J.~Duthu, P.~Garcia, J.~P. Goossens,
  F.~Granet, C.~Grosset-Grange, P.~Guerin, B.~Hebrard, L.~Hilsz, L.~Lamaignere,
  T.~Lacombe, E.~Lavastre, T.~Longhi, J.~Luce, F.~Macias, M.~Mangeant,
  E.~Mazataud, B.~Minou, T.~Morgaint, S.~Noailles, J.~Neauport, P.~Patelli,
  E.~Perrot-Minnot, C.~Present, B.~Remy, C.~Rouyer, N.~Santacreu, M.~Sozet,
  D.~Valla, and F.~Laniesse, \enquote{1.15 {PW}-850 {J} compressed beam
  demonstration using the {PETAL} facility}, Opt. Express 25, 16957--16970
  (2017). DOI: https://doi.org/10.1364/oe.25.016957

\bibitem{Radier2022-ug}
C.~Radier, O.~Chalus, M.~Charbonneau, S.~Thambirajah, G.~Deschamps, S.~David,
  J.~Barbe, E.~Etter, G.~Matras, S.~Ricaud, V.~Leroux, C.~Richard, F.~Lureau,
  A.~Baleanu, R.~Banici, A.~Gradinariu, C.~Caldararu, C.~Capiteanu, A.~Naziru,
  B.~Diaconescu, V.~Iancu, R.~Dabu, D.~Ursescu, I.~Dancus, C.~A. Ur, K.~A.
  Tanaka, and N.~V. Zamfir, \enquote{10 {PW} peak power femtosecond laser
  pulses at {ELI}-{NP}}, High Power Laser Sci. Eng. 10, e21 (2022). DOI:
  https://doi.org/10.1017/hpl.2022.11

\bibitem{Zhu2018-zy}
J.~Zhu, J.~Zhu, X.~Li, B.~Zhu, W.~Ma, X.~Lu, W.~Fan, Z.~Liu, S.~Zhou, G.~Xu,
  G.~Zhang, X.~Xie, L.~Yang, J.~Wang, X.~Ouyang, L.~Wang, D.~Li, P.~Yang,
  Q.~Fan, M.~Sun, C.~Liu, D.~Liu, Y.~Zhang, H.~Tao, M.~Sun, P.~Zhu, B.~Wang,
  Z.~Jiao, L.~Ren, D.~Liu, X.~Jiao, H.~Huang, and Z.~Lin, \enquote{Status and
  development of high-power laser facilities at the {NLHPLP}}, High Power Laser
  Sci. Eng. 6, e55 (2018). DOI: https://doi.org/10.1017/hpl.2018.46

\bibitem{Fuchs2006-zs}
J.~Fuchs, P.~Antici, E.~d'Humières, E.~Lefebvre, M.~Borghesi, E.~Brambrink,
  C.~A. Cecchetti, M.~Kaluza, V.~Malka, M.~Manclossi, S.~Meyroneinc, P.~Mora,
  J.~Schreiber, T.~Toncian, H.~Pépin, and P.~Audebert, \enquote{Laser-driven
  proton scaling laws and new paths towards energy increase}, Nat. Phys. 2,
  48--54 (2006). DOI: https://doi.org/10.1038/nphys199

\bibitem{Wilks1992-yp}
S.~C. Wilks, W.~L. Kruer, M.~Tabak, and A.~B. Langdon, \enquote{Absorption of
  ultra-intense laser pulses}, Phys. Rev. Lett. 69, 1383--1386 (1992). DOI:
  https://doi.org/10.1103/physrevlett.69.1383

\bibitem{Davies2008-ux}
J.~R. Davies, \enquote{Laser absorption by overdense plasmas in the
  relativistic regime}, Plasma Phys. Controlled Fusion 51, 014006 (2008). DOI:
  https://doi.org/10.1088/0741-3335/51/1/014006

\bibitem{Bromage2008-as}
J.~Bromage, S.-W. Bahk, D.~Irwin, J.~Kwiatkowski, A.~Pruyne, M.~Millecchia,
  M.~Moore, and J.~D. Zuegel, \enquote{A focal-spot diagnostic for on-shot
  characterization of high-energy petawatt lasers}, Opt. Express 16,
  16561--16572 (2008). DOI: https://doi.org/10.1364/oe.16.016561

\bibitem{Raffestin2019-pt}
D.~Raffestin, G.~Boutoux, N.~Blanchot, D.~Batani, E.~D'Humi{\`e}res, Q.~Moreno,
  T.~Longhi, H.~Co{\"i}c, F.~Granet, J.~Rault, C.~Liberatore, K.~Jakubowska,
  and V.~Tikhonchuk, \enquote{Application of harmonics imaging to focal spot
  measurements of the `{PETAL}' laser}, J. Appl. Phys. 126, 245902 (2019). DOI:
  https://doi.org/10.1063/1.5129856

\bibitem{Fryxell2000-om}
B.~Fryxell, K.~Olson, P.~Ricker, F.~X. Timmes, M.~Zingale, D.~Q. Lamb,
  P.~MacNeice, R.~Rosner, J.~W. Truran, and H.~Tufo, \enquote{{FLASH: An
  Adaptive Mesh Hydrodynamics Code for Modeling Astrophysical Thermonuclear
  Flashes}}, Astrophys. J. Suppl. Ser. 131, 273--334 (2000). DOI:
  https://doi.org/10.1086/317361

\bibitem{Kruschwitz2012-ls}
B.~E. Kruschwitz, S.-W. Bahk, J.~Bromage, M.~D. Moore, and D.~Irwin,
  \enquote{Accurate target-plane focal-spot characterization in high-energy
  laser systems using phase retrieval}, Opt. Express 20, 20874 (2012). DOI:
  https://doi.org/10.1364/OE.20.020874

\bibitem{Williams2021-ls}
W.~H. Williams, J.~K. Crane, D.~A. Alessi, C.~D. Boley, M.~W. Bowers, A.~D.
  Conder, J.-M.~G. Di~Nicola, P.~Di~Nicola, C.~Haefner, J.~M. Halpin, M.~Y.
  Hamamoto, J.~E. Heebner, M.~R. Hermann, S.~I. Herriot, D.~C. Homoelle, D.~H.
  Kalantar, T.~E. Lanier, K.~N. LaFortune, J.~K. Lawson, R.~R. Lowe-Webb, F.~X.
  Morrissey, H.~Nguyen, C.~D. Orth, L.~J. Pelz, M.~A. Prantil, M.~C. Rushford,
  R.~A. Sacks, J.~T. Salmon, L.~G. Seppala, M.~J. Shaw, R.~J. Sigurdsson, P.~J.
  Wegner, C.~C. Widmayer, S.~T. Yang, and T.~L. Zobrist,
  \enquote{Spatio-temporal focal spot characterization and modeling of the
  {NIF} {ARC} kilojoule picosecond laser}, Appl. Opt. 60, 2288 (2021). DOI:
  https://doi.org/10.1364/AO.416846

\bibitem{Ohland2022-ls}
J.~B. Ohland, U.~Eisenbarth, B.~Zielbauer, Y.~Zobus, D.~Posor, J.~Hornung,
  D.~Reemts, and V.~Bagnoud, \enquote{Ultra-compact post-compressor on-shot
  wavefront measurement for beam correction at {PHELIX}}, High Power Laser Sci.
  Eng. 10, e18 (2022). DOI: https://doi.org/10.1017/hpl.2022.9

\bibitem{Dromey2009-ls}
B.~Dromey, C.~Bellei, D.~Carroll, R.~Clarke, J.~Green, S.~Kar, S.~Kneip,
  K.~Markey, S.~Nagel, L.~Willingale, P.~McKenna, D.~Neely, Z.~Najmudin,
  K.~Krushelnick, P.~Norreys, and M.~Zepf, \enquote{Third harmonic order
  imaging as a focal spot diagnostic for high intensity laser-solid
  interactions}, Laser Part. Beams 27, 243--248 (2009). DOI:
  https://doi.org/10.1017/S0263034609000329

\bibitem{Wang2017-ls}
Z.~Wang, B.~Hu, and Q.~Yin, \enquote{An improved schlieren method for
  measurement and automatic reconstruction of the far-field focal spot}, PLoS
  One 12, e0171415 (2017). DOI: https://doi.org/10.1371/journal.pone.0171415

\end{thebibliography}

\end{document}